\documentclass[twocolumn,trackchanges]{aastex701}

\newcommand\invisiblesection[1]{%
  \refstepcounter{section}%
  \addcontentsline{toc}{section}{\protect\numberline{\thesection}#1}%
  \sectionmark{#1}}

\definecolor{aogblue}{RGB}{43,98,228}

\newcommand{\cosmic}{\textsc{COSMIC}}

\begin{document}

\title{Properties of Core Collapse Supernovae from Binary Population Synthesis}

\author[orcid=0009-0006-3116-7913,sname=Martinez,gname=Mark]{Mark Martinez}
\affiliation{McWilliams Center for Cosmology \& Astrophysics, Department of Physics, Carnegie Mellon University, Pittsburgh, PA 15213, USA}
\email{markmart@andrew.cmu.edu}

\author[0000-0002-7296-6547,gname=Anna,sname=O`Grady]{Anna J. G. O`Grady}
\affil{McWilliams Center for Cosmology \& Astrophysics, Department of Physics, Carnegie Mellon University, Pittsburgh, PA 15213, USA}
\email{aogrady@andrew.cmu.edu}

\author[0000-0001-5228-6598,gname=Katelyn,sname=Breivik]{Katelyn Breivik}
\affil{McWilliams Center for Cosmology \& Astrophysics, Department of Physics, Carnegie Mellon University, Pittsburgh, PA 15213, USA}
\email{kbreivik@andrew.cmu.edu}

\author[0009-0003-5399-3798,gname=Gina,sname=Chen]{Gina Chen}
\affil{McWilliams Center for Cosmology \& Astrophysics, Department of Physics, Carnegie Mellon University, Pittsburgh, PA 15213, USA}
\email{gwchen@andrew.cmu.edu}

\correspondingauthor{Mark Martinez}
\email{markgm@cmu.edu}

\correspondingauthor{Katelyn Breivik}
\email{kbreivik@andrew.cmu.edu}

\begin{abstract}
Core collapse supernovae (CCSNe) impact many areas of astrophysics, including compact object formation and gravitational waves, but many uncertainties remain in our understanding of the evolution of their progenitors. We use the binary population synthesis code \cosmic\ to simulate populations of CCSNe across a wide range of metallicities and binary evolution assumptions. Our models vary the prescriptions for mass transfer stability, common envelope ejection efficiency, natal kick strength, and remnant mass-limited explodability to assess their impact on the resulting population of CCSNe. We find that reproducing the observed Type I to Type II rate requires either low common envelope efficiency or modified prescriptions for common envelope survival, highlighting the importance of stellar mergers in shaping the CCSN population. We further classify our synthetic CCSNe into subtypes and present their relative abundances using several different sets of classification criteria, highlighting the large uncertainties that persist in mapping progenitor properties to spectral classes. Finally, we present delay time distributions (DTDs) for our overall populations, separated into Type I and II, and into the full set of observed subtypes. Our DTDs show that models reproducing the observed Type I to Type II rate produce a larger fraction of late CCSNe than is expected under standard assumptions.
\end{abstract}


\section{Introduction}
Massive stars are the progenitors of core collapse supernovae (CCSNe) and are very likely to be found in binary systems \citep[e.g.][]{Offner23}. The majority of massive stars in binary systems will interact by transferring mass over the course of their stellar lifetimes \citep{Sana12}, which can dramatically shape the properties of the CCSNe this system produces. As a donor star initiates Roche-lobe overflow (RLOF), the mass transfer may remain stable, or become dynamically unstable, resulting in common envelope evolution (CEE), which either leads to a stellar merger or a close binary after a successful ejection of the shared envelope. A donor star that loses most of its hydrogen envelope through mass transfer may produce a Type I CCSN, defined by the complete absence of observed spectral hydrogen lines. In contrast, stellar merger products often result in Type II CCSNe, with visible spectral hydrogen, that create delayed explosions relative to single stars of similar mass \citep{Zapartas2017, Wagg2025}. Asymmetries during core collapse may impart a natal kick to the newly formed neutron star (NS) or black hole (BH). In binary systems disrupted by a natal kick, the companion evolves like a single star, which typically produces Type II CCSNe \citep[e.g.][]{Eldridge2011, Zapartas2017}. If the system remains bound, the resulting orbit becomes eccentric, increasing the likelihood of future mass transfer, and likely CEE if the companion is sufficiently massive. However, despite the importance of CEE and natal kicks to massive CCSN creating binaries, their impact on the resulting CCSN population has not been studied systematically.

Many previous works have used stellar evolution models of single or binary stars, and inferred the properties of the CCSN from the properties of the progenitor star just before core collapse \citep[e.g.][]{Wellstein&Langer1999, Pols&Dewi2002, Heger2003, Eldridge2004, Yoon2010, Claeys2011, Georgy2012, Eldridge2013, Kinugawa2024, Gilkis2025, Souropanis2025}. While our approach follows the same fundamental principle, we extend this framework to explore how the assumptions governing binary evolution shape the resulting CCSN population. In doing so, we introduce several key advances that differentiate our work.

We adopt a mass-dependent binary fraction based on the \textit{multiplicity} fraction of \citet{Offner23}. While many studies assume a fixed binary fraction across mass, we account for the fact that multiplicity rises steeply with stellar mass, and that higher-order systems (quite common at high masses) should be approximated as binaries rather than single stars. As a result, roughly $95\%$ of the CCSNe in our models originate from stars born in binaries. This change substantially impacts the demographics and properties of the CCSNe we produce.

A wide range of key binary interaction parameters is explored, including common envelope efficiency ($\alpha$), mass transfer stability, natal kick dispersion ($\sigma$), and varied maximum allowed remnant mass ($M_{\rm rem}$) below which we consider all CCSNe to produce successful explosions and above which lead to failed, unobservable explosions. Although these parameters are sometimes varied only to check the robustness of results, we treat them as primary levers that directly shape the CCSN population. Our models are also computed across the full metallicity range permitted in \cosmic, allowing us to describe metallicity-dependent behavior which corresponds to a variety of different stellar environments.

Our entire CCSN population is then classified into the commonly observed subtypes (IIP, IIL, IIb, IIn, Ib, Ic), reviewing the observational definitions of these classes, the complications inherent to subtype classification, and the open questions in connecting progenitors at core collapse to observed events. We introduce several classification schemes, each reflecting a different physical or observational bias, and compare how these choices affect subtype demographics and progenitor channels. We make predictions for interaction-powered Type IIn CCSNe, identifying the dominant evolutionary channels that lead to IIn-like circumstellar environments and quantify the fraction of systems that generate dense circumstellar material (CSM) through binary interaction.

We present how our variations in binary evolution physics affect the delay time distributions (DTDs) of our CCSNe. Because DTDs encode the timescales over which stellar populations enrich their host galaxies, they form an essential link between stellar evolution and galactic chemical evolution. We separate the Type I and II populations, and present DTDs resolved by subtype, providing insight into how each class contributes distinct yields by virtue of their ejecta profiles on different timescales.

The structure of this paper is as follows. In Section~\ref{sec:Methods} we describe how we simulate CCSN populations in \cosmic\ and outline our model variations. In Section~\ref{sec:Relative Rates} we examine how our binary-evolution assumptions influence the relative rates of Type I and Type II CCSNe. Section~\ref{Sec:Subtypes} presents our subtype-classification framework, the resulting subtype distributions, and how these depend on the adopted classification scheme. In Section~\ref{Delay times} we present the DTDs of the full, Type I, Type II, and subtype populations for all variations of binary interaction parameters and classification models. Finally, in Section~\ref{sec:Conclusion} we summarize our key findings and highlight their broader implications. All results and code necessary to reproduce this work are publicly available: the full data product is hosted on Zenodo\footnote{\href{https://doi.org/10.5281/zenodo.17620853}{doi.org/10.5281/zenodo.17620853}}, and the analysis toolkit used in this work, available for reuse in other studies, is available on GitHub\footnote{\href{https://github.com/MarkGM02/COSMIC-ccsnlab}{github.com/MarkGM02/COSMIC-ccsnlab}}.

\section{Simulating Binary CCSN Progenitors} \label{sec:Methods}
We use \cosmic\footnote{\href{https://cosmic-popsynth.github.io}{cosmic-popsynth.github.io}}, a binary population synthesis suite, to quantify the effects of different assumptions for binary-star interactions on the properties of CCSNe. \cosmic\ implements single star evolution with SSE \citep{SSE}, commonly known as the Hurley tracks. The binary star evolution algorithm is based on BSE \citep{BSE} and includes several updates to the treatment of binary interactions and compact object formation. See \citet{Breivik2020} for a detailed description of these updates.

\subsection{Initializing stellar and binary populations} \label{Sampling Parameters}
To simulate CCSN populations, we consider both single and binary star progenitors where the population size is fixed to contain $\sim10^6$ binaries\footnote{The actual population size varies by statistical noise from sampling the initial mass function.}. The number of single stars is determined by the mass-dependent binary fraction which we implement following the the multiplicity fraction reported in \citet{Offner23}.

All simulated populations are initialized with the same initial mass function and binary parameter distributions. Our initial condition sampling begins with random draws from an initial mass function following \citet{Kroupa01}; we then probabilistically determine which masses are assigned binary companions based on the mass-dependent binary fraction. All binary systems are then assigned secondary masses with a uniform mass ratio, such that the minimum secondary mass is defined by the pre-main-sequence contraction lifetime of the primary star\footnote{This is specified as \texttt{qmin=-1} in \cosmic}. Orbital periods and eccentricities for each binary are drawn according to the prescriptions defined in \cite{Sana12}. \cosmic\ assigns minimum orbital periods such that the primary star fills less than half of its Roche volume and a fixed maximum orbital period of $10^{5.5}\,\rm{days}$. This differs from the \citet{Sana12} maximum period of $\sim3000\,\rm{days}$.

\cosmic\ computes single star evolution as a function of metallicity, where this is defined relative to solar metallicity. In this study, we set solar metallicity to $Z_\odot=0.02$, which is consistent with the stellar evolution tracks used in SSE \citep{Pols98}.

\subsection{Assumptions for Single and Binary Star Evolution}
In all of our simulations, we apply a consistent wind prescription that accounts for the strong metallicity dependence of line-driven stellar winds on the MS and post-MS \citep[e.g.][]{Vink2001, Vink2011, Meynet&Maeder2005, Grafener&Hamman2008}. We first apply the wind mass loss prescriptions of \citet{Nieuwenhuijzen+1990:1990A&A...231..134N} for massive MS stars with $L>4000\,L_{\odot}$ and \citet{Kudritzki+1978:1978A&A....70..227K} for giant stars. For O and B stars, we use the metallicity-dependent wind models of \citet{Vink2001}, extended to apply across temperature ranges $12{,}500\,\mathrm{K} < T_\mathrm{eff} < 25{,}000\,\mathrm{K}$ and $25{,}000\,\mathrm{K} < T_\mathrm{eff} < 50{,}000\,\mathrm{K}$ separately, following the approach of \citet{Dominik2013} and \citet{Rodriguez2016}. Winds from Wolf–Rayet stars are modeled using the prescription of \citet{Vink&deKoter2005}. Finally, for stars beyond the Humphreys-Davidson limit we apply a fixed wind mass loss rate of $10^{-4}\,M_{\odot}\,\rm{yr}^{-1}$ to mimic the effect of luminous blue variable (LBV) eruptions rather than modeling them explicitly.

We explore the effect of CEE and compact-object natal kicks on the properties of CCSNe by varying our assumptions for how they are treated during binary evolution. Specifically, we vary: i) the mass transfer stability criteria that dictate whether a system undergoing RLOF enters into CEE (q$_{\rm crit}$), ii) the CEE envelope ejection efficiency ($\alpha$), and iii) the characteristic dispersion for assigning natal kick velocities ($\sigma$). We discuss how \cosmic\ treats CEE and natal kicks in Sections~\ref{Common Envelope Variations} and \ref{Natal Kick Variations} respectively. All other prescriptions for how binary-star interactions proceed follow the default settings in \cosmic.

\subsubsection{Common Envelope Evolution} \label{Common Envelope Variations}
\cosmic\ determines the stability of mass transfer using critical mass ratios which account for how the donor responds to mass loss and how the orbit responds to mass exchange as described in \citet{BSE}. When RLOF begins a CEE is initiated if the prescribed critical mass ratio is exceeded, such that $M_\mathrm{don}/M_\mathrm{acc} > q_\mathrm{crit}$. We adopt two prescriptions for determining this critical mass ratio. 

Our fiducial model\footnote{This is specified as \texttt{qcflag = 5} in \cosmic} follows \citet{Neijssel2019} where core hydrogen burning stars are assigned $q_\mathrm{crit}=1.717$, and stars crossing the Hertzsprung Gap are assigned $q_\mathrm{crit}=3.825$ based on models in \citet{Vigna-Gomez2018}. Fits from \citet{Hjellming&Webbink1987} based on condensed polytropes for deeply convective stars are applied to more evolved hydrogen-rich stars. Mass transfer from stripped envelope stars is always assumed to be dynamically stable ($q_\mathrm{crit}=\infty$) motivated by \citet{Tauris2015, Tauris2017} and \citet{Vigna-Gomez2018} which showed that this assumption is necessary to replicate the observed galactic BNS population. Finally, white-dwarf donors are assigned $q_\mathrm{crit}=0.628$, following \citet{BSE}.

We also consider a model variation that follows Section 5.1 of \citet{Belczynski2008}\footnote{This is specified as \texttt{qcflag = 4} in \cosmic} where hydrogen rich stars with both radiative and convective envelopes have $q_\mathrm{crit}=3$. Stripped-envelope helium MS stars are assigned $q_\mathrm{crit}=1.7$ and evolved helium stars are assigned $q_\mathrm{crit}=3.5$ following \citet{Ivanova2003}. White dwarf donors are limited to $q_\mathrm{crit}=0.628$, following the prescription of \citet{BSE}. Finally, we note that our application of fixed mass transfer stability for phases of evolution rather than as a function is a significant simplification \citep[e.g.][]{Ge+2010:2010ApJ...717..724G, Ge+2015:2015ApJ...812...40G}.

We model CEE using the $\alpha\lambda$ formalism where orbital energy is used to unbind the shared envelope \citep{Webbink+1984:1984ApJ...277..355W}
\begin{equation}
    \frac{E_\mathrm{bind,grav}}{\lambda} \leq \alpha E_\mathrm{orb}.
\end{equation}
The common envelope efficiency parameter ($\alpha$) scales how much orbital energy is contributed to the ejection of the envelope and $\lambda$ sets the binding energy of the envelope to its stellar core. The envelope is successfully ejected when the envelope binding energy is surpassed by the injected energy. For a review of CEE see \citet{Ivanova2013}.

CEE is highly uncertain. Several studies have aimed to constrain $\lambda$ and $\alpha$ using binaries hosting low-mass progenitors of white dwarfs \citep[e.g.][]{Zorotovic+2010:2010A&A...520A..86Z,DeMarco+2011:2011MNRAS.411.2277D,Zorotovic+2014:2014A&A...568A..68Z, Scherbak+2023:2023MNRAS.518.3966S, Yamaguchi+2024:2024MNRAS.52711719Y} and higher mass progenitors to compact objects like NSs and BHs \citep[e.g.][]{Dominik+2012:2012ApJ...759...52D,Kruckow+2016:2016A&A...596A..58K, Giacobbo+2018:2018MNRAS.480.2011G, Fragos+2019:2019ApJ...883L..45F,Marchant+2021:2021A&A...650A.107M}
To capture this uncertainty, we vary $\alpha$ between 0.05 and 5. We fix $\lambda$ to follow the prescription in the appendix of \citet{Claeys2014} which defines fits to the envelope density profiles of the stars in the \citet{Pols98} models. We assume that no ionization energy is used to eject the envelope, though models with $\alpha>1$ imply an unknown additional energy source which could be attributed to ionization energy.

In addition to explicitly varying $\alpha$, we apply envelope ejection criteria based on \textsc{MESA} simulations from \citet{Klencki2021}. These criteria are based on envelope ejection outcomes from a CEE event with a massive giant donor star and a BH companion across metallicities ($0.01Z_\odot$ to $Z_\odot$) as functions of i) the ZAMS mass of the donor and ii) the radius of the donor at the onset of RLOF. The first prescription uses $\alpha=1$ and fixed mass transfer stability criteria for radiative and convective donors such that $q_\mathrm{crit,rad}=1.5$ and $q_\mathrm{crit,conv}=3.5$. The second uses $\alpha=0.7$ and a smooth transition between $q_\mathrm{crit}$ for convective and radiative donors.

To apply these criteria, we evolve a binary in \cosmic\ until it reaches a CEE event. We select the model from \citet{Klencki2021} with the nearest metallicity to the current population in log space. We evaluate whether the radius of the donor at the start of RLOF is larger than the threshold needed for survival, which is a function of the ZAMS mass of the donor. If the donor radius falls below this threshold, the evolution is resumed with $\alpha\approx0$ to ensure a stellar merger results. If this threshold is exceeded, evolution resumes with either $\alpha=1.0$ or $0.7$ and the value of $\lambda$ from the appendix. We then continue the CEE, and the outcome is determined by the balance between envelope binding energy and orbital energy.

In addition to the $\alpha$ variations and \citet{Klencki2021} models, we consider two more test cases. One with $\alpha=1$ and pessimistic CEE that assumes that all CEE events involving donors without a well-defined core–envelope boundary result in a merger \citep{Belczynski2008, Ivanova2008}. This includes all MS stars, stars crossing the Hertzsprung Gap, and white dwarfs. We lastly consider a variation in which all CEE onsets result in a merger.

\subsubsection{Compact Object Formation and Natal Kicks} \label{Natal Kick Variations} 
For stellar progenitors which reach core collapse, we assign remnant masses with the analytic prescription for a delayed-convection explosion presented in \citet{Fryer2012}. This approach fills the mass gap between NSs and BHs. We set a maximum NS mass of $3\,M_\odot$, above which the remnants are classified as BHs. We further assign natal kicks randomly by drawing kick velocities from a Maxwellian distribution with a velocity dispersion $\sigma$ then reducing the strength of the natal kick based on the fraction of mass that falls back onto the proto compact object following \citet{Fryer2012}. We vary $\sigma$ between $50\,$km\,s$^{-1}$ and $265\,$km\,s$^{-1}$ to capture the range of natal kick strengths that have been inferred from both isolated neutron stars and neutron stars in binaries \citep[e.g.][]{Hobbs2005, Igoshev+2021:2021MNRAS.508.3345I,Valli+2025:2025arXiv250508857V}. BH natal kicks are fallback-modulated following \citet{Fryer2012}. All natal kick directions are sampled isotropically.

Several works have suggested that stars with main-sequence masses around $\sim8$–$11M_\odot$ may end their lives without forming an iron core. Instead, these stars develop sufficiently large helium cores and undergo core collapse triggered by electron capture on $^{24}$Mg and $^{20}$Ne, producing an electron capture supernova (ECSN) \citep[e.g.][]{Miyaji1980, Nomoto1984, Nomoto1987, Podsiadlowski2004, Ivanova2008, Poelarends08}. We set the range of potential helium core masses to create an ECSN between $1.6M_\odot$ and $2.25M_\odot$, consistent with \citet{BSE}. We also allow ECSNe to be triggered through binary interactions. If an oxygen-neon white dwarf accretes material from a binary companion and exceeds $1.38\,M_{\odot}$, an accretion-induced collapse will occur. Likewise, a merger or collision between two carbon-oxygen or oxygen-neon white dwarfs results in a merger-induced collapse when the total mass exceeds the $1.38\,M_{\odot}$ threshold. Since an ECSN is expected to occur through a rapid explosion rather than a delayed neutrino-driven explosion, studies have argued for smaller NS natal kicks relative to typical CCSNe \citep{Podsiadlowski2004, Ivanova2008}. Therefore, we draw the natal kick velocities for ECSNe from a Maxwellian distribution with $\sigma_\mathrm{ECSN} = 20$km\,s$^{-1}$ in all our simulations.

Ultra-stripped supernovae (USSNe) occur when a stripped star, having undergone RLOF with a compact object companion, proceeds to core collapse. Motivated by their low ejecta masses \citep{Tauris2013, Tauris2015}, we similarly adopt a lower characteristic kick dispersion for USSNe, drawing their natal kick velocities from a Maxwellian distribution with $\sigma_\mathrm{USSN} = 20$km\,s$^{-1}$.

\subsection{Binary Evolution Grids} \label{Binary evolution variations}

\begin{deluxetable*}{llll}
\label{tab:cosmic_params}
\tablehead{
\colhead{Parameter} & \colhead{Symbol} & \colhead{Fiducial Model} & \colhead{Variations}
}
\startdata
\texttt{metallicity} & $Z$ & 0.020243 & [0.0001 ... 0.03] \\
\texttt{qcflag} & --- & 5 & [4 5] \\
\texttt{sigma} & $\sigma$ & 265.0 km\,s$^{-1}$ & [50.0 ... 265.0] km\,s$^{-1}$ \\
\texttt{alpha1} & $\alpha$ & 1.0 & [0.05 ... 5.0] \\
\texttt{sigmadiv}\tablenotemark{a} & $\sigma_{\mathrm{ECSN}}, \sigma_{\mathrm{USSN}}$ & 20 km\,s$^{-1}$ & --- \\
\enddata
\caption{A list of the key evolution parameters in this study, including \cosmic\ code flags and their corresponding physical variable names.}
\tablenotetext{a}{\cosmic\ requires \texttt{sigmadiv} = $-20.0$ to directly set $\sigma_{\mathrm{ECSN}}, \sigma_{\mathrm{USSN}}$ to 20 km\,s$^{-1}$.}
\end{deluxetable*}

We construct two primary variation grids by holding one of the $\alpha$–$\sigma$ pair fixed while varying the other. In the first grid, we fix $\alpha$ at its fiducial value of $1.0$ and vary $\sigma$ over ten linearly spaced values between $50\,$km\,s$^{-1}$ and $265\,$km\,s$^{-1}$. In the second grid, we fix $\sigma$ to $265\,$km\,s$^{-1}$ and vary $\alpha$ across ten logarithmically spaced values between $0.05$ and $5.0$. For each $\alpha$–$\sigma$ combination, we apply both mass transfer stability prescriptions and evolve 30 initial stellar populations with logarithmically spaced metallicities from $Z = 0.0001$ ($0.005Z_\odot$) to $Z = 0.03$ ($1.5Z_\odot$). The list of varied parameters is summarized in Table~\ref{tab:cosmic_params}.

\section{Relative Rates of Type I and Type II CCSNe} \label{sec:Relative Rates}

We first consider how assumptions for binary interactions alter the relative rates of Type I and Type II CCSNe by assigning each star that undergoes core collapse (including USSN progenitors) to one of these two classes. We exclude ECSNe, pulsational pair-instability supernovae, pair-instability supernovae, and any supernovae that do not leave behind a compact remnant. Although some work suggests that hydrogen ejecta masses as low as $M_{\rm ej,H} \approx 0.001\,M_\odot$ may suffice for spectroscopic identification as Type II \citep[e.g.][]{Dessart2011}, we adopt a higher threshold, requiring $M_{\rm ej,H} \geq 0.033\,M_\odot$, in agreement with \citet[][]{Hachinger2012, Gilkis2022}. All CCSNe with ejecta masses below this threshold are assigned to be Type I.

\begin{figure*}
    \centering
    \includegraphics[width=\textwidth]{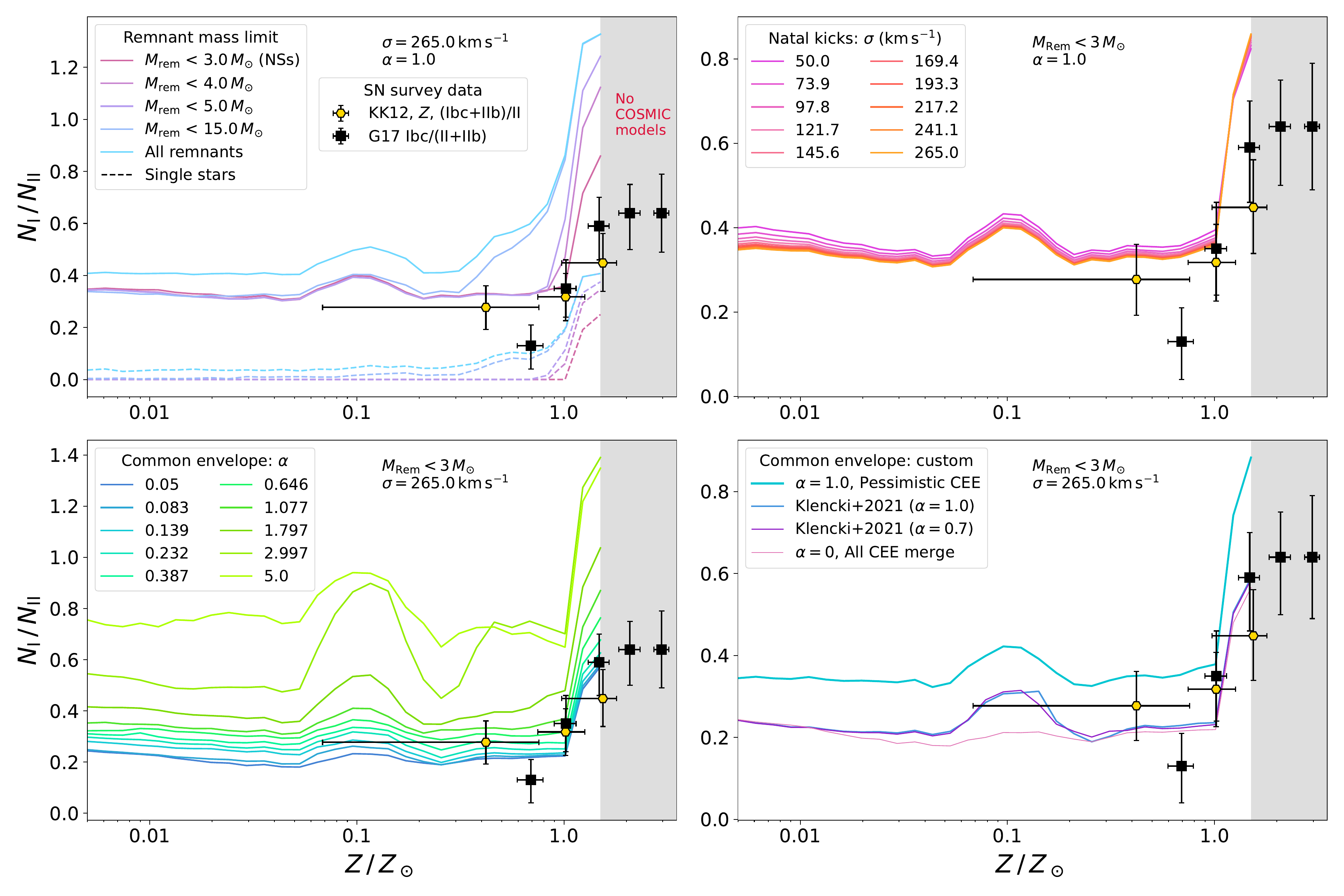}
    \caption{Ratio of Type I to Type II CCSNe across the metallicity range of our models. Each panel shows a different set of model variations: the maximum allowed remnant mass for single (dashed) and binary-inclusive (solid) populations (top left), the natal kick velocity dispersion $\sigma$ (top right), the common-envelope efficiency $\alpha$ (bottom left), and custom prescriptions for common-envelope evolution (bottom right). Data points from two surveys are shown for comparison: the LOSS survey as reported in \citet{LOSS} (black squares) and the sample from \citet{KK12} (yellow hexagons). Both datasets are measured on the T04 metallicity scale \citep{T04}, which we convert to $Z$ using the solar reference value of  $12 + {\rm log(O/H)} = 9.0$. Grey shaded regions at $Z > 1.5\,Z_\odot$ mark metallicities beyond the upper limit of our \cosmic\ models.}
    \label{fig:Loss Plots}
\end{figure*}

We primarily compare our findings with the data from the Lick Observatory Supernova Search (LOSS) survey \citep{Li2011}, particularly Figure 10 of \citet{LOSS}. The LOSS survey normalized the detection of SNe within target galaxies through the use of control times, which reduced the bias against the lower absolute magnitudes of certain subtypes. In Figure~\ref{fig:Loss Plots} we compare our \cosmic\ simulations against the data from LOSS and the sample from \citet{KK12}, assembled from the Asiago catalog \citep{Barbon1999} and the Palomar Transient Factory \citep{Arcavi2010}.

Both datasets have metallicities reported on the T04 metallicity scale \citep{T04}. In the LOSS survey, these metallicities correspond to central galaxy abundances, and the authors of \citet{LOSS} argue that these values should approximate the metallicities at the locations of their CCSNe to within $\lesssim0.1$ dex. The metallicities in \citet{KK12}, however, were measured as close to the explosion sites as possible. The BPASS v2 models presented in \citet{LOSS} convert their stellar metallicities $Z$ to the T04 scale using a solar abundance reference of $12+\log({\rm O/H})\approx9.0$, since their models at $2Z_\odot$ reach $12+\log({\rm O/H})\approx9.3$. They note that the systematic uncertainties in comparing metallicities derived from observational galaxy measurements to those inferred from stellar models are of order $0.1$ dex (see \citet{Modjaz2011, Anderson2015} for a review). The canonical solar reference value presented in \citet{Asplund2009} is $12+\log({\rm O/H})=8.69$. 

Considering the combined effects of i) explosion-site versus central abundance offsets, ii) the T04 calibration’s tendency to slightly overestimate metallicities compared to other scales \citep{Kewley&Ellison2008}, and iii) other cumulative systematic uncertainties, we adopt $12+\log({\rm O/H})=9.0$ as an effective solar reference for converting these datasets into $Z$. This choice yields excellent agreement between the data and our models, as it aligns the metallicity at which stellar winds produce a sharp increase in the relative rate of Type I CCSNe.

\subsection{Explodability Criteria}

The upper left panel of Figure~\ref{fig:Loss Plots} shows our fiducial assumptions with $\sigma=265.0\,{\rm km\,s}^{-1}$ and $\alpha=1.0$. Each model corresponds to a varied maximum allowed remnant mass $M_{\rm rem}$, below which we consider all CCSNe to produce successful explosions and above which lead to failed, unobservable explosions. Models that include only single stars are shown with dotted lines, and models including binaries are shown with solid lines. We find that single stars can nearly reproduce the observed rate of Type I to II CCSNe when we consider CCSNe producing any remnant mass. Still, the simulated ratio of Type I to II is broadly lower than the observed ratio and appears to asymptote to too low of a rate, $N_{\rm I}/N_{\rm II} \approx 0.4$, whereas the data from \citet{LOSS} suggests the end behavior should be nearer to $N_{\rm I}/N_{\rm II} \approx 0.6$. We further note that the highest remnant masses ($\gtrsim20\,M_{\odot}$) are expected to lead to direct collapse where no explosion is observable. Thus, single star populations are unable to faithfully match the observed rates.

When including binary stars in the population, we find that this model over predicts the relative number of Type I CCSNe across metallicities regardless of our limit for $M_{\rm rem}$. This over prediction is especially high at the upper end of our metallicity range, where we find that this model can produce up to $N_{\rm I}/N_{\rm II} \approx 1.2$ when all remnants are considered. 

We find that decreasing the limit for $M_{\rm rem}$ consistently decreases the relative rate of Type I CCSNe. In our models, the proportion of Type II CCSNe which form massive remnants ($M_{\rm rem}>3.0\,M_\odot$) is relatively constant and a bit lower than the proportion of Type I CCSNe below $Z/Z_\odot\approx0.3$, about $30\%$ and $46\%$ respectively. At $Z/Z_\odot\approx0.3$, these proportions diverge drastically, and our Type II CCSNe drop to a minimum $\approx0.2\%$ forming massive remnants, and our Type I CCSNe reach a maximum of $\approx59\%$. Type I CCSNe are $\approx1.5$ times more likely than type II CCSNe to form massive remnants at low metallicities, and up to $\approx220$ times more likely at our highest metallicity. Because of this drastic trend, in order to match observations, we find that we must tighten our explodability criteria to only allow remnants $<3\,M_\odot$. We maintain this assumption for the other panels.

In several of our models (and all panels of Figure~\ref{fig:Loss Plots}) a ``bump'' appears, centered around $Z/Z_\odot\approx0.1$, which is a nonphysical artifact of the SSE fitting formulae which reach maximum radii in this metallicity range. This radius increase corresponds directly to an increase in the number of systems that undergo RLOF and lose their hydrogen envelopes. This increased interaction causes the feature to be strongly pronounced in the binary inclusive population, while there is a much smaller effect which appears in our single star populations. The corresponding increase in luminosity with larger radii also increases the strength of stellar winds. These winds are, in most cases, not strong enough to remove the hydrogen envelope entirely; however, they can in a small number of cases. We
discuss the most prominent impact of these increased stellar winds later in Section \ref{Sec:Subtypes}.

\subsection{Natal kicks}

We show the impact of varying the natal kick velocity dispersion $\sigma$ in the upper right panel of Figure~\ref{fig:Loss Plots}. We find that the impact of this variation is modest because the only changes in outcome occur for the binaries which have two stellar components that are massive enough to undergo core collapse, restricting the overall impact to a small fraction of the total stellar population. Increasing natal kicks causes the binary to disrupt more frequently and causes the stellar companion to undergo the rest of its evolution as a single star. As shown in our single-star models discussed above, this most commonly produces a Type II CCSN except in the rare cases where the envelope is stripped due to high stellar winds. Binary systems that remain bound after a natal kick generally have eccentric orbits, which increase the likelihood of interaction in the future due to a reduction in the periapse distance where we expect RLOF to occur. If the secondary has sufficiently high mass to produce a CCSN, it is very likely to be more massive than its compact object companion\footnote{This is especially true for the case where we restrict to $M_{\rm{rem}}<3\,M_{\odot}.$}. This large mass ratio leads to high rates of CEE during the second RLOF in our models for binaries which remain bound after the first CCSN. Because the donor stars are massive, these CEE frequently result in a merger between the star and its compact object companion. This interaction is assumed to completely disrupt the star in COSMIC but could also lead to the formation of Thorne-$\dot{Z}$ytkow objects \citep{Thorne+1977:1977ApJ...212..832T}. However, in either case, we do not expect the secondary to produce a standard CCSN.

Generally, the binary systems that are disrupted by natal kicks are biased towards being wide at the time of core collapse. This can occur either because the binary is initially wide enough to avoid a first RLOF or because the initially less massive star gains a significant amount of mass during the first RLOF such that a mass ratio reversal widens the binary. In the case where the secondary does not gain a significant amount of mass, it is expected to produce a Type II CCSN since it will not have its envelope stripped. In the case of a mass ratio reversal, there is a threshold mass and metallicity at which the secondary's wind mass loss is strong enough to strip its own envelope. These combined effects lead to metallicity-specific trends in both the absolute and relative rates of Type I and Type II CCNSe. At low metallicities, we find that binaries disrupted due to higher natal kicks exclusively increase the number of Type II CCSNe in the population. This effect persists until $Z\approx\,Z_\odot$, at which point the rates of both Type I and II CCSNe increase by similar amounts because self-stripping through winds leads to more Type I CCSNe.

\subsection{Common Envelope Evolution}
\label{Impact of Common Envelope}

The lower two panels of Figure~\ref{fig:Loss Plots} show our variations for the treatment of CEE. We find that $N_{\rm I}/N_{\rm II}$ varies dramatically across our models. CEE influences the properties of CCSN produced by both stellar components in a given binary. A successful CEE ejection completely removes the donor envelope, leading to a Type I CCSN. Since a successful CEE ejection extracts orbital energy from the binary the resulting binary separation is very close. Therefore, the binary is more likely to survive the first CCSN and natal kick, often leading to another CEE, which produces a second Type I CCSN upon ejection, or none after a merger that fully disrupts the secondary.

Decreasing $\alpha$ strongly increases the rate of stellar mergers that occur in either the first or second RLOF. Hydrogen rich merger products undergo the rest of their evolution as single stars and usually produce Type II CCSNe. We find that decreasing $\alpha$ in our models decreases the ratio $N_{\rm I}/N_{\rm II}$ at nearly every metallicity, with the only exception being two of our high alpha models overlapping at super solar metallicities. At these high metallicities, very massive hydrogen rich merger products may be sufficiently massive such that winds strip their hydrogen envelope, making merger products produce Type I instead of Type II CCSNe.

For our two highest $\alpha$ values, we find a metallicity-specific trend where $\alpha=5$ produces fewer Type I CCSNe than $\alpha\approx3$. This is because binaries that survive the first CEE have wider orbits with increasing $\alpha$ and at higher metallicities, larger wind mass loss rates serve to widen the binary even more. The effect of this relative widening diminishes for the $\alpha=5$ because the rate of binary disruptions increases with wider orbits.

We find that the models following \citet{Klencki2021}, shown in the bottom right panel of Figure~\ref{fig:Loss Plots}, have the strongest agreement with the G17 and KK12 data. We further find that simply enforcing our pessimistic CEE assumption, in which all stars without a well-defined core-envelope boundary merge,for models with $\alpha=1$, over predicts the $N_{\rm I}/N_{\rm II}$ ratio at near-solar metallicities. For models where we assume that all binaries undergoing CEE will merge, we find a broad agreement with both \citet{Klencki2021} models. Based on the close agreement between the \citet{Klencki2021} model with $\alpha=0.7$ and \citet{LOSS, KK12}, we select this model for future discussion of the subtypes within Type I and Type II in Sections~\ref{Sec:Subtypes} and \ref{Delay times}.

\subsection{Comparison to Previous Results}
We compare our results in Figure~\ref{fig:Loss Plots} to two previous works which produced a similar figure. The first is in \citet{LOSS} which included simulations from \textsc{BPASS V2} \citep{Eldridge2016, Stanway2016}. We find similar agreement that reproducing the observed $N_{\rm I}/N_{\rm II}$ ratio at low metallicities is only possible when binary populations are considered.
We further find that the simulated ratio of $N_{\rm I}/N_{\rm II}$ from \textsc{BPASS} generally agrees with our metallicity specific trends, including a significant increase in the rate of Type I CCSNe toward near-solar metallicity. We do note, however that the increase in the ratio occurs near $0.5\,Z_\odot$, and more gradually for the \textsc{BPASS} models while we find the increase to occur sharply at $Z_\odot$. Finally, we find that our single–star models generally produce fewer Type I CCSNe than the \textsc{BPASS} models at low metallicities, but that our increase in Type I CCSNe from single stars is more abrupt at higher metallicities, and after this sharp rise, our simulated $N_{\rm I}/N_{\rm II}$ exceeds the \textsc{BPASS} predictions. This may be due to our application of high wind mass loss for stars that are subject to the Humphreys-Davidson Limit.

We also find general agreement with the \textsc{POSYDON} models of \citet{Souropanis2025} which show a similar behavior of a flat ratio of $N_{\rm I}/N_{\rm II}$ at sub-solar metallicities between $0.2-0.4$. Comparing directly to their Ibc/II model, we find a sharper increase in the ratio for $Z/Z_{\odot}\geq1$. A moderate increase in the ratio agrees more strongly with the Ibc/II ratio from \cite{KK12}. However, the rapid rise near $2Z_{\odot}$ exhibited in \citet{LOSS} agrees more with our simulations.

The ratio $N_{\rm I}/N_{\rm II}$ is defined slightly differently across these studies. In \citet{LOSS}, the denominator (``II'') includes all Type II CCSNe, with IIb events counted among them. The numerator therefore contains only the Type I CCSNe. In contrast, the dataset from \citet{KK12} excludes IIb events from their denominator and instead assigns them to the numerator along with the Type I CCSNe. It is not entirely clear how the \textsc{BPASS} models in \citet{LOSS} handled this classification, but in the Ibc/II models of \citet{Souropanis2025}, IIb SNe are excluded from the denominator. The authors of \citet{LOSS} note that shifting IIb events between numerator and denominator has only a minor effect on the ratio. The same is true for our models, as we show in the next section, we produce very few IIb CCSNe overall. In contrast, the models of \citet{Souropanis2025} produce many more IIb events, a consequence of differences in how mass transfer is treated in \textsc{POSYDON}.

Our models with $\alpha>1$ predict a much higher ratio of $N_{\rm I}/N_{\rm II}$ when comparing to both the \textsc{BPASS} and \textsc{POSYDON} models. This is due to successful CEE ejections combined with the higher binary fraction used in our simulations that follows the the mass-dependent multiplicity fraction of \citet{Offner23}. Under these assumptions, stars of $8\,M_{\odot}$ have a binary fraction of about $92\%$, which rises with mass and reaches a maximum of roughly $96\%$ at higher masses. Conversely, mass-independent binary fractions are used in the \textsc{BPASS} models of \citet{LOSS}, where half of the CCSN population is produced in binaries, and in the \textsc{POSYDON} models of \citet{Souropanis2025} where the binary fraction is $60\%$.

\section{CCSN Subtypes and Classification}
\label{Sec:Subtypes}
In this section, we present the subtype demographics of our Type I and Type II populations. We discuss the challenges involved in connecting the outcomes of our models to observed subtypes and outline the rationale behind our classification schemes. We then demonstrate how these schemes influence the relative proportions of subtypes and the trends in their formation channels. We also examine how our adopted explodability criteria affect these results.

\subsection{Type II Subtypes}

CCSNe that retain most or all of their hydrogen envelopes span a diverse set of observational outcomes. Large observed samples of Type II CCSNe suggest that Types IIb, IIL, and IIP do not represent strict, discrete classes, but a continuum \citep[e.g.][]{Anderson2014, Sanders2015, Galbany2016, Valenti2016} in which the hydrogen envelope mass at core collapse drives the diversity in observed features \citep[e.g.][]{Nomoto1995, Heger2003, Dessart2011, Eldridge2018, Hiramatsu2021}. Type IIn CCSNe are identified by narrow emission lines in their spectra which are caused by interaction with nearby circumstellar material. These features often occur from flash ionization, in which emission lines may disappear within hours or days \citep[e.g.][]{Yaron2017, Bullivant2018, Bruch2021, Jacobson-Galan2022, Terreran2022}. If observations are made only at later times, these signatures can be missed.

Mounting evidence also suggests that some Type IIL and IIP CCSNe also experience varying levels of circumstellar interaction \citep[e.g.][]{Moriya2011, Khazov2016, Hosseinzadeh2018, Jacobson-Galan2022, Jacobson-Galan2024, Jacobson-Galan2025}, further blurring subtype boundaries. In practice, separating these subclasses requires assumptions for how interaction-powered events should be treated and how the hydrogen envelope mass should be mapped onto photometric categories. To address this ambiguity, we adopt two classification schemes that differ in how they prioritize IIn versus IIP progenitors. We first describe the details of these schemes and then present the resulting subtype demographics across metallicity.

\subsubsection{Classification Method and Variations}
\label{Classification method and variations II}

To form our classifications, we rely on two key progenitor properties: i) the maximum mass-loss rate in the last kyr before core collapse ($\dot{M}_{\rm max,kyr}$), and ii) the hydrogen ejecta mass $M_{\rm ej,H}$. We describe the justification and treatment of these as follows.

\smallskip \noindent \textbf{Predicting CSM interaction:} X-ray observations suggest that Type IIn progenitors must exhibit mass-loss rates of $\gtrsim10^{-3}\,M_\odot\,\mathrm{yr}^{-1}$ to explain observed CSM densities \citep[][]{Chandra2012a, Chandra2012b, Dwarkadas&Gruzko2012, Katsuda2014}. Light-curve modeling of interaction-powered SNe similarly supports high mass-loss rates in the final decades before explosion \citep[][]{Moriya2013, Moriya2014, Ransome&Villar2025}.

Many observed Type IIn CCSNe are consistent with progenitors in an LBV-like phase \citep[e.g.][]{Gal-Yam&Leonard2009, Smith2011, Mauerhan2013, Ransome&Villar2025}. Line-driven LBV winds typically range from $\sim10^{-5}$-$10^{-4}\,M_\odot\,\mathrm{yr}^{-1}$ \citep[][]{Vink&deKoter2002, Vink&deKoter2005, Smith2004}, but the extreme mass-loss rates creating Type IIn CCSNe likely arise from brief eruptive phases reaching (or surpassing) $\sim10^{-3}\,M_\odot\,\mathrm{yr}^{-1}$ \citep[e.g.][]{Humphreys&Davidson1994, Hillier2001, Smith2011LBV}. These eruptions are not modeled in \cosmic; instead, stars beyond the Humphreys-Davidson limit have mass loss rates of $10^{-4}\,M_\odot\,\mathrm{yr}^{-1}$ and this rate is not dynamically increased prior to core collapse. 

Binary interaction has also been proposed to explain some interacting SNe \citep{Kashi&Soker2010, Smith2024, Ransome&Villar2025} through mechanisms such as mergers during CEE or collisions involving compact objects, explored in many works \citep[e.g.][]{Kashi&Soker2006, Langer2012, Smith2018, Schroder2020}. Because of this, we indiscriminately use the mass loss rate as our proxy, which includes binary interaction as well as stellar winds as the underlying cause.

To identify interaction-powered events we flag progenitors with $\dot{M}_{\rm max,kyr} \geq 10^{-4}\,M_\odot\,\mathrm{yr}^{-1}$. This conservative threshold corresponds to the high end of LBV winds and allows us to capture progenitors that might plausibly power CSM interaction, even if eruptive events are not explicitly modeled. This approach is consistent with work showing pre-SN mass loss in IIn events increases in the decades before explosion \citep[][]{Moriya2014, Ransome&Villar2025}. 

\smallskip \noindent \textbf{Using hydrogen mass to distinguish Type IIb/L/P:} Previous studies have proposed $M_{\rm ej,H}\sim0.5-0.6\,M_\odot$ as the boundary between SNe IIb and IIL \citep{Podsiadlowski1993, Woosley1994, Claeys2011}, and $M_{\rm ej,H}\sim1.5$–$2\,M_\odot$ for the IIL/IIP boundary \citep{Eldridge2004, Georgy2012, Heger2003}. We instead adopt the framework of \citet{Hiramatsu2021}, which systematically modeled light curves across a grid of hydrogen masses, explosion energies, and $^{56}$Ni yields. They identified envelope-mass transitions corresponding to approximate photometric boundaries. We include what they identified as short-plateau SNe II into the IIL class. Following them, we adopt $M_{\rm ej,H} = 0.91\,M_\odot$ for the IIb/L boundary and $M_{\rm ej,H} = 4.5\,M_\odot$ for the IIL/P boundary.

Using this framework we consider two classification schemes that differ primarily in whether CSM interaction or massive hydrogen ejecta are prioritized first in the decision tree.

\smallskip \noindent \textit{\textbf{Branch IIn first:}} In this scheme, any progenitor with evidence for strong pre-SN mass loss is first classified as Type IIn. Remaining Type II events are divided by $M_{\rm ej,H}$ into IIb, IIL, and IIP.  

\smallskip \noindent \textit{\textbf{Branch IIP first:}} In this alternate scheme, progenitors with large hydrogen envelopes are first classified as Type IIP. Next, interaction-powered progenitors are identified as IIn. The remainder are divided into IIb and IIL by $M_{\rm ej,H}$.

\smallskip These two schemes capture the plausible range of subtype demographics, reflecting both early-time classification ambiguities and the fact that massive hydrogen envelopes can produce plateau light curves even with CSM interaction.

\subsubsection{Subtype Properties}

\begin{figure*}
    \centering
    \includegraphics[width=\textwidth]{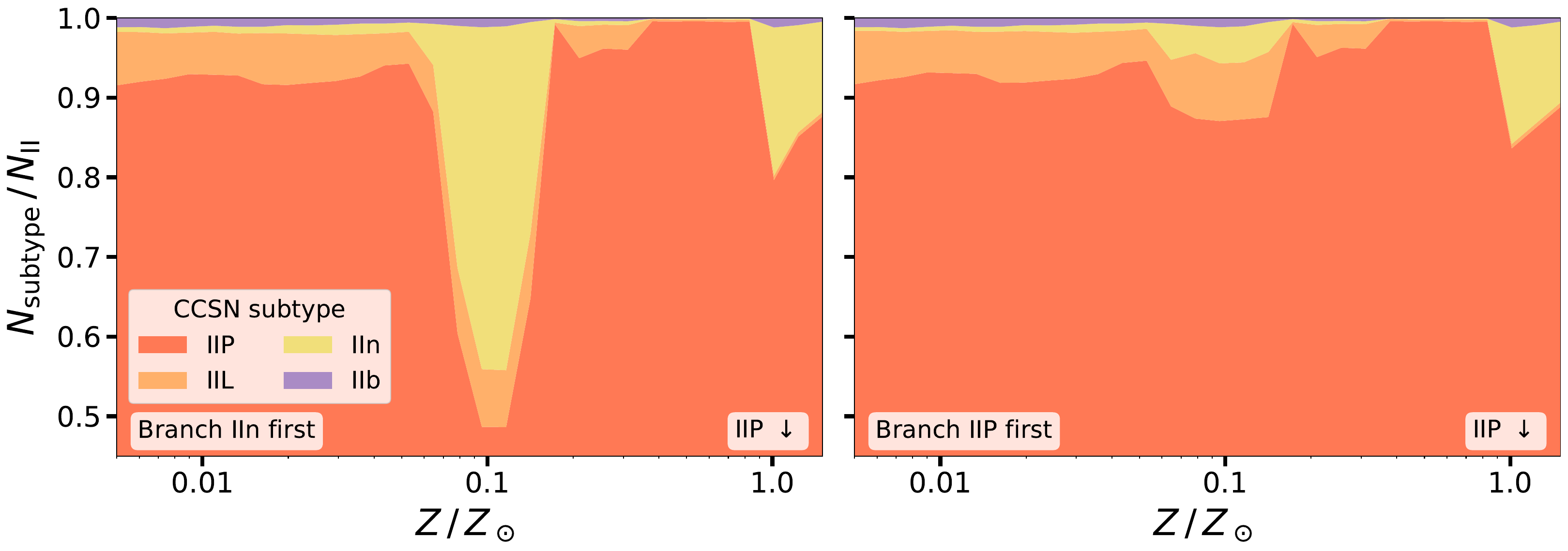}
    \caption{Stacked subtype fractions of Type II CCSNe across metallicity. The model shown follows the CEE prescriptions of \citet{Klencki2021}, and only includes CCSNe forming small remnants ($M_{\rm rem}<3.0\,M_\odot$). The left and right panels apply the \textit{``branch IIn first''} and \textit{``branch IIP first''} classification schemes respectively (see Section~\ref{Classification method and variations II}). Subtypes are colored as shown in the legend and stacked such that their sum is unity. The lower bound of the plot is raised to aid in viewing, since Type IIP CCSNe dominate the population.}
    \label{fig:subtypes_II}
\end{figure*}

In Figure~\ref{fig:subtypes_II} we present the demographics of our CCSNe according to the two classification schemes described above, including only CCSNe forming remnants $<3\,M_\odot$. The primary impact of changing schemes lies in the $Z/Z_\odot\sim0.1$ centered ``bump". In this metallicity range, the exchange is (by construction) only between Type IIn and IIP CCSNe. This reflects the fact that many progenitors in this metallicity band retain $\geq4.5\,M_\odot$ of hydrogen while also experiencing substantial pre-explosion mass loss. In practice, this implies that if all CSM interaction were detectable, these events might appear as IIn at early times, but be classified as IIP if observed only later. We present the results of note by subtype next.

\smallskip \noindent \textbf{Type IIP:} The progenitors of Type IIP CCSNe remain the most uniform across metallicity, typically accounting for about $90\%$ of the Type II population, despite briefly dipping to about $47\%$ at intermediate metallicities in the \textit{``branch IIn first''} scheme (left panel of Figure~\ref{fig:subtypes_II}). This decrease is largely mitigated in the \textit{``branch IIP first''} scheme (right panel of Figure~\ref{fig:subtypes_II}), where IIP events constitute at least $87\%$ of the population in this metallicity band. However, changing between classification schemes does not substantially impact the bulk properties of IIP progenitors. Among Type II CCSNe, IIP progenitors span the widest range of ZAMS masses. Across metallicity, the lower mass limit is steady at $\sim1\,M_\odot$ in merging systems, where a secondary engulfs a more massive primary after a mass-ratio-reversal. The upper limit increases modestly with metallicity, averaging around $18\,M_\odot$ but reaching nearly $20\,M_\odot$ at solar metallicity. The relative contributions of the effectively single-star and merger channels vary substantially over metallicity. Most IIP progenitors are not the initially more massive star; over our metallicity grid the share of progenitors originating from primaries grows slightly from $\sim37$-$46\%$ with increasing metallicity.

Across all metallicities, between $\sim$68–75\% of IIP progenitors experience some form of binary interaction, with this fraction generally increasing with metallicity. This interaction trend reflects two opposing sub-trends over our metallicity grid. The contribution from binaries that experienced only stable mass transfer declines from $\sim26$–$18\%$, while the fraction that underwent CEE increases steadily from $\sim43$–$57\%$. Given that a successful CEE ejection strips the donor’s envelope completely, progenitor systems undergoing CEE necessarily produce stellar mergers. Accordingly, the fraction of IIP CCSNe from merger products increases from $\sim45$–$72\%$ with increasing metallicity. While IIP events remain the dominant subtype at all metallicities, and remain approximately constant in proportion to the number of Type II CCSNe, stellar mergers from lower mass progenitors are required to contribute at higher metallicities due to increasing winds.

\smallskip \noindent \textbf{Type IIn:} The progenitors of Type IIn CCSNe span nearly the same ZAMS mass range as IIP progenitors, with a lower bound set by merger products and an slightly higher upper limit at most metallicities of $\sim$19–21$\,M_\odot$. However, their demographic and evolutionary properties are far more sensitive to metallicity and to the branching scheme—than any other Type II subtype.

A defining feature of the Type IIn population is the prominent intermediate-metallicity ``bump’’ near $Z/Z_\odot\sim0.1$–0.3, coinciding with the regime in which stellar radii fits from \citet{SSE} reach their maximum values. Because the wind-mass-loss models we use scale with stellar temperature and therefore with radius at fixed luminosity, progenitors in this band experience dramatically enhanced wind mass loss during the last thousand years before collapse. As a consequence, the IIn fraction jumps to nearly $43\%$ in the \textit{``branch IIn first’’} scheme. In contrast, in the \textit{``branch IIP first’’} scheme, many of these events are recategorized as IIP, lowering the IIn fraction to only $\sim5\%$.

Outside this intermediate-metallicity regime, Type IIn progenitors separate into two broad behaviors. At low metallicity ($Z/Z_\odot \lesssim 0.05$), IIn CCSNe constitute only $\sim1\%$ of the Type II population. in this case, stellar winds are comparatively weak, and as a result roughly half of low-metallicity IIn progenitors exhibit wind-driven mass loss in the final thousand years, with the remainder undergoing mass transfer.
At solar and super-solar metallicity, the IIn fraction rises again. In the “branch IIP first’’ scheme, IIn events make up $\sim15\%$ of Type II CCSNe around $Z\sim Z_\odot$, increasing to $\sim18\%$ in the \textit{``branch IIn first’’} scheme. Nearly all high-metallicity progenitors experience strong winds, powering the CSM in up to $\sim98\%$ of cases at the highest metallicities.

Our results can be directly compared to the \textsc{MESA} binary models of \citet{Ercolino+2025:2025arXiv251004872E}, who identify IIn progenitors exclusively as systems that undergo mass transfer after core helium depletion. They report IIn fractions of roughly 2–4\% of the total core collapse population, which corresponds to about 5–7\% of their Type II CCSNe. Since they only considered models at solar metallicity, this compares directly to our 15-18\% range. One possible reason for our larger IIn fraction is the relatively larger radii that stars can have in COSMIC via the fits from \cite{SSE} relative to the radii in \textsc{MESA}. Because our wind-mass-loss models scale with larger radii as described above, our simulations likely include an enhanced IIn fraction when compared with \citet{Ercolino+2025:2025arXiv251004872E}.

\smallskip \noindent \textbf{Type IIL:} As shown in Figure \ref{fig:subtypes_II}, Type IIL CCSNe typically make up the next largest contribution, and their frequency declines with increasing metallicity. At most metallicities, their progenitors occupy a mass range comparable to that of IIP progenitors, with an average upper limit near $18\,M_\odot$ and a slightly higher maximum around $21\,M_\odot$. Their lower limit is modestly elevated, at roughly $2\,M_\odot$, still arising from mergers in which a secondary engulfs a more massive primary, though the minimum mass required for this pathway is somewhat higher. The vast majority of these progenitors come from three distinct channels.

For $Z/Z_\odot\lesssim0.3$, about $80\%$ of these progenitors come from a stellar merger produced by a failed CEE event initiated by the secondary. These systems are low mass, with neither star massive enough to form an iron core on its own. They undergo mass-ratio-reversal after a phase of stable mass transfer in which the primary donates its entire hydrogen envelope to the secondary and eventually becomes a white dwarf. The secondary later fills its Roche lobe, initiates CEE, and the system merges. About $7\%$ originate from non-interacting primaries, and the remaining $\sim13\%$ come from primaries that lose only part of their envelope through stable mass transfer.

Above $Z/Z_\odot\sim0.3$, more than $99.1\%$ of IIL progenitors interact before core collapse. The contribution from the merger channel described above declines steadily, reaching about $13\%$, while the contribution from the stable mass-transfer channel rises to roughly $87\%$.

\smallskip \noindent \textbf{Type IIb:} Type IIb CCSNe are rare in our models, much rarer than observations suggest \citep{Shivvers2017}. Their frequency appears to be nearly constant throughout metallicity, contributing about $1\%$ of the type II population. This fraction slightly decreases to a minimum of $\sim0.1\%$ near solar metallicity, before returning to a similar level, and is unaffected by the classification choice. The scarcity of IIb events in our simulations highlights the difficulty in reproducing observed rates without more detailed treatments of hydrogen envelope stripping, since COSMIC does not allow for partial envelope removal except in the case of stars which settle onto the `red clump' of the Hertzsprung-Russell Diagram after ascending the first giant branch.

A small fraction of our IIb progenitors originate from stellar mergers, typically mass-ratio–reversed systems, in which the primary loses its entire hydrogen envelope, and a low-mass secondary engulfs the helium-rich primary. This produces a merger product that ultimately reaches core collapse with only a few tenths of a solar mass of hydrogen. These systems, however, account for less than about $13\%$ of the cumulative total across metallicity. The dominant pathway, comprising of $\sim85\%$ of IIb progenitors, consists of primaries with initial masses on the higher end of the Type II range. For these primary-origin IIb progenitors, the median ZAMS mass increases with metallicity, rising from roughly $10\,M_\odot$ to about $20\,M_\odot$ across our metallicity grid. Up to roughly solar metallicity, every single progenitor donates its hydrogen envelopes through stable mass transfer, leaving behind only a few tenths of a solar mass. The remaining cumulative $2\%$ of progenitors come from a channel present only at $Z/Z_\odot > 1$. At these metallicities, winds become strong enough that some IIb progenitors are effectively single, with winds alone stripping the envelope down to the same small hydrogen layer produced by stable mass transfer at lower metallicities. This channel can account for as many as $35\%$ of the IIb population at a given super-solar metallicity, yet contributes only about $2\%$ of Type IIb CCSNe across our metallicity grid.

\subsection{Type I Subtypes}

Observationally delineating Type Ib from Type Ic CCSNe is subject to several complications, including but not limited to the timing and wavelength regime considered \citep[e.g.][]{Dessart2012, Dessart2020, Lu2025}, and the effects of line blending and signal-to-noise limitations, which can obscure or mimic the presence of key He I features \citep[e.g.][]{Dessart2015}. Similar to these observational challenges, directly connecting observed Type Ib and Ic CCSNe to their progenitors is complicated by degeneracies in simulated light curves and spectra that can be produced through different assumptions for explosion energy, $^{56}$Ni yield, and the spatial distribution (mixing) of $^{56}$Ni throughout the ejecta even if the same ejecta profile is used \citep[e.g.][]{Teffs2020, Lu2025}. Although we are limited by these uncertainties, we next present several ways in which we classify our progenitor models, with justifications that aim to target different sources of uncertainty in the progenitor–explosion connection.

\subsubsection{Classification Method and Variations}
\label{Classification method and variations I}

In order to create our classification schemes, we consider observable quantities and properties of the progenitor star and its ejecta. Specifically, we consider the total ejecta mass ($M_{\rm ej}$), the helium ejecta mass ($M_{\rm ej,He}$), and whether the progenitor donates mass to a companion while it is already a helium star. We classify all Type I CCSNe as either Ib or Ic. We note that applying the same mass-loss criteria to attempt to distinguish Type Ibn CCSNe (a reasonable assumption given their similarly predicted mass-loss rates \citep{Maeda22}) yields a small number of candidates, of order $0.1\%$ of Type I CCSNe. We therefore choose to separate the population only into Ib and Ic events and reserve a more detailed exploration of Ibn progenitors for future work. We consider three schemes for separating Type Ib and Ic CCSNe, which are as follows:

\smallskip \noindent \textit{\textbf{Absolute helium:}} Our first classification scheme is based on the idea that some threshold helium ejecta mass sets the boundary between Type Ib and Ic CCSNe, under the expectation that sufficient helium will generate the necessary He I spectral lines observed for Ib events. This absolute helium criterion is the most common approach in the literature. Early population studies \citep[e.g.][]{Wellstein&Langer1999, Pols&Dewi2002} adopted a dividing line of $M_{\rm ej,He}=0.5\,M_\odot$, while \citet{Georgy2009} instead applied $M_{\rm ej,He}=0.6\,M_\odot$ to capture a limiting case where the cutoff significantly altered progenitor mass ranges. These studies, which largely predated detailed radiative transfer modeling, made approximate but self-consistent choices. Already, however, the complications were clear: \citet{Wellstein&Langer1999} emphasized the role of nickel mixing in shaping helium line visibility, and \citet{Filippenko&Barth&Matheson1995} showed that helium can be spectroscopically detected in some events otherwise classified as Type Ic, suggesting only a very small amount of helium may be sufficient to leave an imprint.

Radiative transfer work since then has revealed that the physical picture is far more nuanced, defining a ``hidden helium'' problem. \citet{Hachinger2012} demonstrated that the spectral characteristics of He can persist below $M_{\rm ej,He}\sim0.14\,M_\odot$, while subsequent studies have found that the effective limit strongly depends on the distribution of $^{56}$Ni and the degree of mixing. For example, \citet{Dessart2015} showed that up to $0.3\,M_\odot$ of helium can be hidden when more carbon-oxygen material is mixed outward. Models from \citet{Teffs2020} show that the He I 1.083~$\mu$m line can saturate with as little as $0.02\,M_\odot$. Similarly, \citet{Williamson2021} argued for an upper limit of $0.05\,M_\odot$ based on optical He~I lines in SN 1994I. Recently, the models of \citet{Lu2025} demonstrated visible He features with just $M_{\rm ej,He}=0.02\,M_\odot$, and emphasized that the strength of helium features is not directly proportional to helium mass, but rather controlled by the interplay of the radiation field and the composition of the ejecta.

In light of this diversity of results, there is no clear consensus on the exact helium mass dividing Type Ib and Ic. We primarily adopt the $M_{\rm ej,He}=0.14\,M_\odot$ threshold from \citet{Hachinger2012}, as a practical median choice. However, we also discuss the impact of varying this threshold below.

\smallskip \noindent \textit{\textbf{Relative helium:}} Our second scheme follows the approach first proposed in \citet{Eldridge2011}, which argues that the ratio of helium ejecta mass to total ejecta mass provides a better proxy for observability than an absolute helium mass threshold. In their work, this ratio was calibrated to $M_{\rm ej,He}/M_{\rm ej}=0.56$ to reproduce the sample from \citet{Smartt2009}. The same method was later adopted by \citet{Eldridge2013}, who adjusted the ratio to 0.61 to match their updated sample. The underlying physical motivation of this choice is that very small amounts of nearly pure helium ejecta should be observable as Type Ib. If the ejecta is instead dominated carbon or oxygen, particularly if the helium is well mixed, helium spectral features could be suppressed such that a Type Ic is observed. These limiting cases are not well captured by the \textit{``absolute helium''} approach. Furthermore, \citet{Dessart2020} argues that the mass fraction of helium in the ejecta is fundamentally linked to the processes by which helium is excited, since the decay power and associated non-thermal rates scale with this mass fraction. We discuss our choice for the fixed ratio below.

\smallskip \noindent \textit{\textbf{Mass transfer:}} Our final classification scheme identifies any star that donates mass after already becoming a stripped helium star as a Type Ic progenitor \citep{Tauris2013, Tauris2015}. The motivation for this approach is that rapid population synthesis codes like COSMIC may not provide sufficiently detailed descriptions of stellar envelope structure, since they rely on analytic prescriptions to approximate mass transfer and stripping processes. As a result, explicitly tracking whether a system experienced mass transfer may serve as a better indicator of envelope stripping than relying exclusively on the final envelope mass.

\subsubsection{Subtype Properties}

\begin{figure*}
    \centering
    \includegraphics[width=\textwidth]{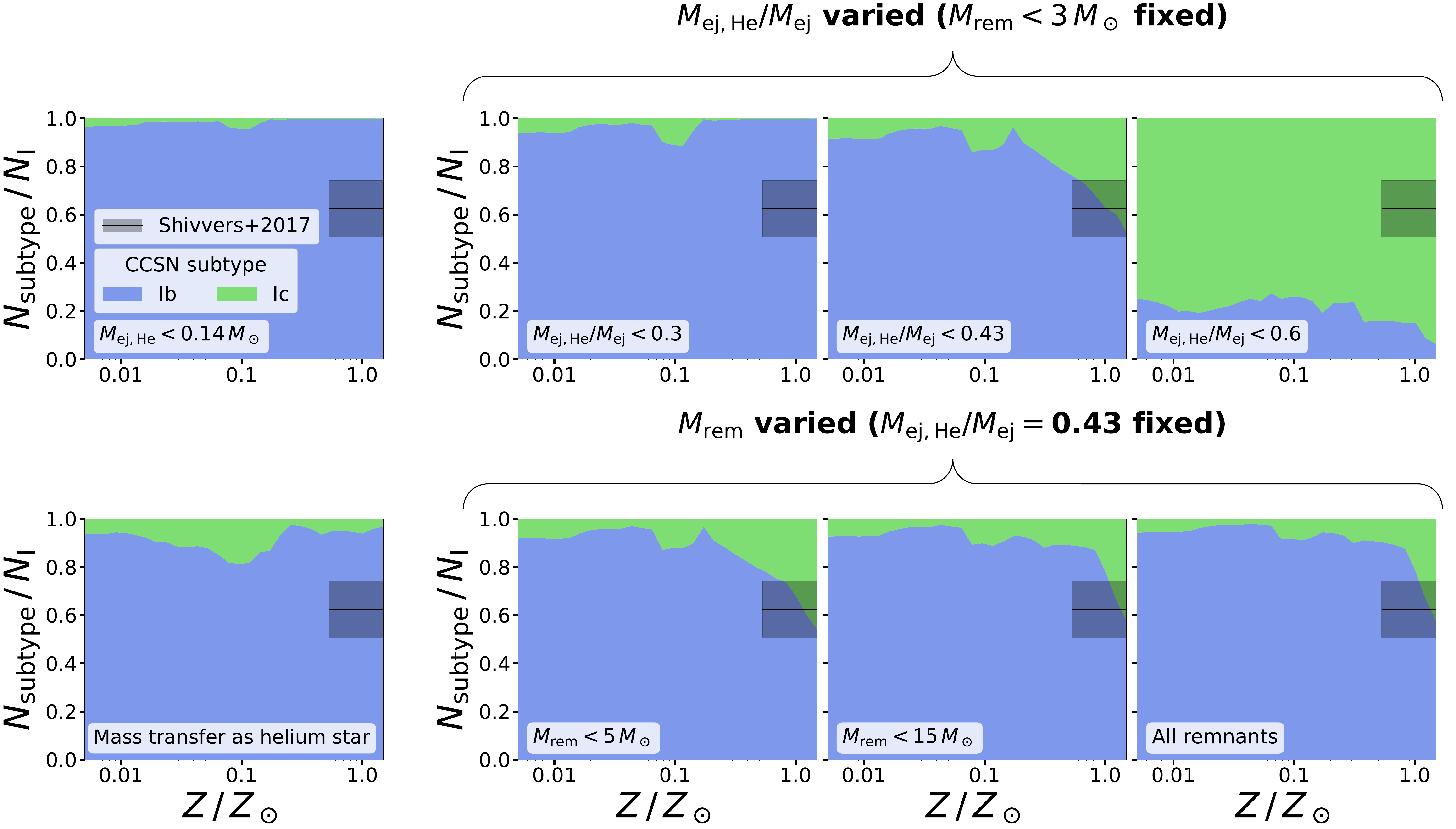}
    \caption{Stacked subtype fractions of Type I CCSNe across metallicity. The model shown follows the CEE prescriptions of \citet{Klencki2021}, and only includes CCSNe forming small compact object remnants ($M_{\rm rem}<3.0\,M_\odot$) for all panels other than the lower right series. Each panel includes the observed Ib fraction from \citet{Shivvers2017}, shown as a flat line across the metallicity range of the LOSS survey, with shaded regions indicating the associated uncertainty (the high-metallicity end extends beyond our models and is not shown, see Section~\ref{sec:Relative Rates} for the treatment of metallicity comparison). The upper and lower left panels show the \textit{``absolute helium''} and \textit{``mass transfer''} classification schemes, respectively (see Section~\ref{Classification method and variations I}). The upper right three panels show a sequence varying the ratio $M_{\rm ej,He}/M_{\rm ej}$, which defines classification in the \textit{``relative helium''} scheme. The lower right sequence maintains the relative helium threshold of $M_{\rm ej,He}/M_{\rm ej}=0.43$ while varying the allowed maximum remnant mass.}
    \label{fig:subtypes_I}
\end{figure*}

Since we find our simulated Type I CCSNe to be commonly associated with the production of larger stellar remnants, we also discuss the variation of a maximum threshold for $M_{\rm rem}$ below which we consider CCSNe to be successful. In Figure~\ref{fig:subtypes_I} we present the subtype abundances corresponding to each classification scheme, including the results of varying the ratio $M_{\rm ej,He}/M_{\rm ej}$ and exploring different maximum allowed remnant masses when using this ratio-based approach. Figure~\ref{fig:subtypes_I} shows each of our classification schemes, as well as the effects of varying the remnant mass limit, in comparison to the Ib/Ic rate computed in \citet{Shivvers2017}. The results from each series of panels are as follows.

\smallskip \noindent \textit{\textbf{Absolute helium:}} The upper left panel shows that the \textit{``absolute helium''} scheme with $M_{\rm ej,He}=0.14\,M_\odot$ produces very few Ic systems. In fact, even raising our absolute helium threshold to $M_{\rm ej,He}=0.5\,M_\odot$ hardly has an impact, since most systems eject around a solar mass of helium. Not until we raise the threshold to $M_{\rm ej,He}\approx1.2\,M_\odot$ do we match the observed ratio of observed Ic CCSNe in \citet{Shivvers2017}. We regard such a high threshold as physically unjustified. Although our standard choice underproduces Type Ic CCSNe, we continue to adopt $M_{\rm ej,He}=0.14\,M_\odot$ threshold in the discussion that follows.

We find that Ic formation requires binary interaction across all metallicities, whereas Ib systems may arise from non-interacting wide binaries or single stars at solar and super-solar metallicities. Still, even at super-solar metallicities, the vast majority of the Type Ib CCSNe classified using this scheme originate from interacting binaries ($\gtrsim85\%$). In this scheme, Ib and Ic progenitors share similar lower ZAMS-mass limits. These lower limits remain below the minimum ZAMS mass for a single star to produce a CCSN at all metallicities, and correspond to a combination of mergers and accretors that grow sufficiently massive to experience strong stellar winds, often followed by later phases of mass transfer.

However, Ib progenitors extend to significantly higher masses at all metallicities. The most massive Ic progenitors reach only about $16\,M_\odot$ at any metallicity. The maximum mass of Ib progenitors increases with metallicity, rising from roughly $40\,M_\odot$ at the lowest metallicities in our grid to just above $80\,M_\odot$ at the highest metallicities. This trend occurs because we restrict successful explosions to remnants with $M_{\rm rem}<3\,M_\odot$, and stronger winds at higher metallicity reduce the final core mass and thus remnant mass. We find that these very massive stars do lose a substantial fraction of their helium envelopes through winds; however, they still retain more helium than the $0.14\,M_\odot$ threshold required by this scheme.

In our models, the ratio of Type Ib to Ic CCSNe remains roughly constant at low metallicity. This ratio surprisingly decreases at higher metallicity. The leading expectation is that the fraction of Type Ic relative to Type Ib CCSNe should rise with metallicity, as binary interaction strips the hydrogen envelope at a similar rate regardless of metallicity, and metallicity-dependent winds remove the remaining helium layer, forming more Type Ic CCSNe at higher metallicities \citep{Fang2019, Sun2023}. Although recent work suggests that helium-star winds may be weaker than previously assumed \citep{Drout2023, Drout2023MC}, our models fail to produce any Type Ic CCSNe through helium star winds\footnote{Which use the \citet{Hamann&Koesterke1998} Wolf-Rayet wind prescription reduced by a factor of ten following \citet{Yoon&Langer2005}, with a metallicity dependence from \citet{Vink&deKoter2005}.}. Rather, all helium stars in our models require a phase of mass transfer to sufficiently reduce their envelope below $0.14\,M_\odot$ and form a Type Ic CCSN.

\smallskip \noindent \textit{\textbf{Mass transfer:}} The \textit{``mass transfer''} scheme produces a systematically higher number of Type Ic CCSNe than the \textit{``absolute helium''} scheme at all metallicities. In this scheme, the most massive stars at ZAMS still result in Type Ib CCSNe. Ic progenitor masses are similar regardless of metallicity, with an upper limit of $21\,M_\odot$ on average, slightly higher than the upper range in the \textit{``absolute helium''} scheme. For higher metallicities, $Z/Z_\odot$ A pronounced peak in the Ic fraction appears at the metallicities corresponding to the aforementioned maximum-stellar-radius ``bump,'' which is expected since this scheme explicitly requires binary interaction for Ic formation. However, the overall abundance of Ic systems still appears underestimated relative to observations, as the maximum value of $\sim20\%$ of Type Ic CCSNe is lower than that predicted by \citet{Shivvers2017}. 

Some Type I CCSNe classified as Ib in the \textit{``absolute helium''} scheme become Ic under the \textit{``mass transfer''} scheme, signaling that these progenitors interacted as helium stars. It is predominantly these systems that produce the observed peak at intermediate metallicity, indicating that some progenitors may donate mass as helium stars yet still retain a few tenths of a solar mass of helium in our models. While this scheme again fails to reproduce the full observed trends in the Type I CCSN population, it does show that binary interaction involving helium star progenitors can contribute a non-trivial portion of the population, suggesting that traditional explanations relying solely on single helium star evolution may be insufficient.

\smallskip \noindent \textit{\textbf{Relative helium:}} In the three panels in the top left of Figure~\ref{fig:subtypes_I}, we apply the \textit{``relative helium''} scheme with varying ratio $M_{\rm ej,he}/M_{\rm ej}$ to separate Ib from Ic systems. We find that varying this ratio leads to very substantial differences in our results. With the ratio at 0.3, this effectively reduces to the absolute scheme, with very slight variations. When we reach 0.43, the number of Ic systems is dramatically different, and now shows a clearly increasing rate across metallicity, matching both the expected behavior, and the proportion reported in \citet{Shivvers2017}. Further increasing the ratio to 0.6, Ic CCSNe become dominant at every metallicity, and are more common than observed. This ratio is extremely sensitive around $M_{\rm ej,he}/M_{\rm ej}=0.45$, since this is at the center of the peak of the $M_{\rm ej,he}/M_{\rm ej}$ distribution for our Type I CCSNe. Most Type I CCSNe are found to have $M_{\rm ej,he}/M_{\rm ej}$ between 0.4 and 0.8.

Adopting the ratio $M_{\rm ej,he}/M_{\rm ej}=0.43$ matches both the expected behavior and value of the Ic fraction best. As noted in \citet{Eldridge2013}, we note that using this approach creates a more overlapping continuum of progenitor masses. For $Z/Z_\odot<0.1$, Ic progenitors are all $\lesssim20\,M_\odot$, while Ib progenitors extend to higher masses. For metallicities $Z/Z_\odot>0.1$, Ib and Ic progenitors occupy effectively the same initial range of progenitor masses, with Ic CCSNe actually extending to the highest metallicity at super-solar metallicities. The small population of non-interacting progenitors is now nearly evenly divided between Ib and Ic progenitors, making up no more than a few percent of progenitors of either subtype at a given super-solar metallicity.

Since $M_{\rm ej,he}/M_{\rm ej}=0.43$ is broadly successful in reproducing observations, we enforce this criteria in the three lower right panels of Figure~\ref{fig:subtypes_I}, which show variations in the maximum allowed $M_{\rm rem}$ below which CCSNe are considered successful. We find that increasing $M_{\rm rem}$ increases the proportion of Ib CCSNe at all metallicities. When allowing successful CCSNe to leave behind remnants more massive than $3\,M_\odot$, Ib progenitors now extend to higher initial masses at all metallicities.

\section{Delay Time Distributions}
\label{Delay times}

In this section we present the DTDs produced by our suite of COSMIC simulations. We first present the DTDs of our entire population, divided into Type I and Type II CCSNe to discuss metallicity, compact object formation, and binary evolution. We conclude by presenting the DTDs divided by subtype, showing first how our classification variations affect these results, and second how the DTDs of these subtypes correspond to the ZAMS mass of the progenitor, the progenitor mass at core collapse, and the mass of the compact object formed.

\subsection{Remnant Formed and Metallicity}

\begin{figure*}
    \centering
    \includegraphics[width=\textwidth]{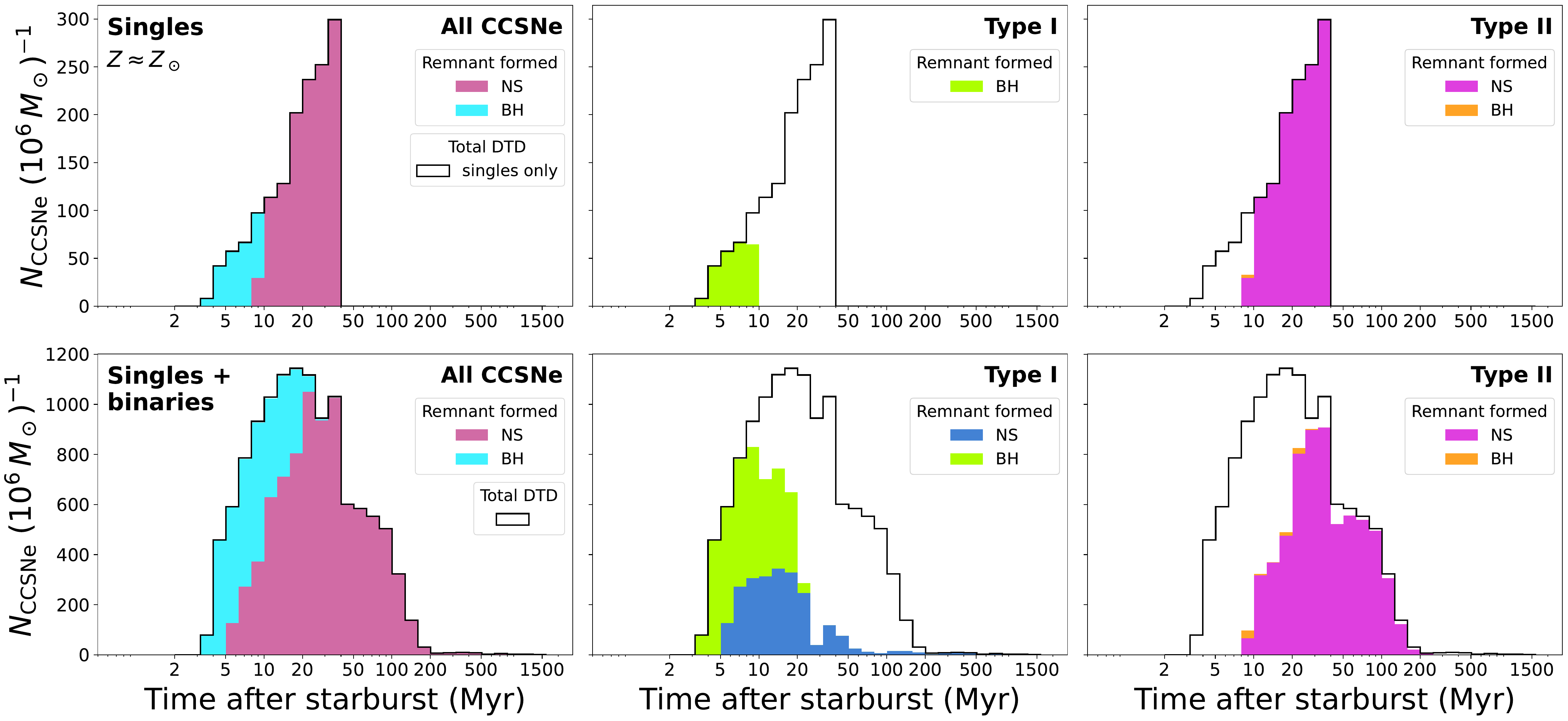}
    \caption{DTDs showing the remnant types of single-only and binary-inclusive populations. The model shown follows our fiducial binary evolution assumptions ($\sigma=265\,{\rm km\,s}^{-1}$, $\alpha=1$) and has metallicity $Z\approx Z_\odot$. The three panels along the top show only single stars, while the lower panels include both single and binary systems. Columns correspond to all CCSNe (left), Type I only (middle), and Type II only (right). Black outlines show the total DTD in each panel. Stacked histograms separate contributions from NSs ($M_{\rm rem}<3\,M_\odot$) and BHs ($3\,M_\odot<M_{\rm rem}<15\,M_\odot$).}
    \label{fig:DTDs_Remnant}
\end{figure*}

In Figure~\ref{fig:DTDs_Remnant} we present DTDs showing single stars along the top three panels, and our population including binaries in the lower three. This population is evolved at solar metallicity, and the fiducial assumptions for binary interaction ($\sigma=265.0\,{\rm km\,s}^{-1}$, $\alpha=1.0$). We use a stacked histogram to separate the contributions of CCSNe that form NSs ($M_{\rm rem}<3\,M_\odot$) and those that form BHs (limited here with $M_{\rm rem}<15\,M_\odot$). The columns in each row, from left to right show, i) all CCSNe, ii) only Type I CCSNe, and iii) only Type II CCSNe.

We find, similarly to \citet{Zapartas2017}, that including binary systems substantially increases the proportion of ``late'' CCSNe. These result predominately from stellar mergers prior to the second CCSN and are overwhelmingly Type II. We similarly find comparable numbers of CCSNe produced per solar mass sampled as \citet{Zapartas2017}. We find a much smaller number of CCSNe per mass sampled in our population of single stars, but this corresponds to the different ways our samples are constructed. Our entire population is sampled according to the \citet{Kroupa01} IMF, and then a mass dependent binary fraction is applied to assign companions. Since this binary fraction increases at higher masses, the lower mass part of the sample is heavily biased toward single stars, meaning that the majority of the stars not massive enough to create CCSNe are single, and do not contribute to this rate. 

We find that BH-forming CCSNe are generally found to occur before NS-forming CCSNe due to their more-massive progenitors evolving more quickly. The population of BH-forming CCSNe reaches a maximum around $\sim8\,$Myr in both the single and binary inclusive population. NS-forming CCSNe reach a later maximum, with two peaks around $\sim25\,{\rm Myr}$ and $\sim35\,{\rm Myr}$. When we decompose the overall population into Type I and II, we see that this first peak corresponds to a maximum in the NS forming Type I population, and the near maximum Type II population, and this later peak corresponds to the true peak of the Type II population.

We again see that massive remnants are associated with Type I CCSNe. In fact we see here that in the single star population, type I CCSNe \textit{only} form BHs, and $\lesssim0.25\%$ of Type II CCSNe form BHs. In the population dominated by binaries, we find that Type I CCSNe now form BHs and NSs in similar numbers, while Type II CCSNe form slightly more BHs, but still form NSs $>99.7\%$ of the time. The small number of Type II CCSNe forming BHs in our models correspond to systems that begin with an initial mass of $\sim21$-$22\,M_\odot$ as single or non-interacting stars, or begin slightly below and accrete up to a similar mass.

The very latest CCSNe in our models are, intriguingly, all Type I. These arise from an exotic evolutionary pathway involving two survived phases of unstable mass transfer, between an initially $\approx6.5$–$7.5\,M_\odot$ primary and an initially $\approx2.2$–$2.4\,M_\odot$ secondary, which ultimately produces a close oxygen-neon white dwarf–helium star binary that merges. The systems in this channel can reach delay times as long as $\sim1550,{\rm Myr}$. Observational DTD studies typically treat times $\gtrsim400\,{\rm Myr}$ as the domain of prompt Type Ia supernovae \citep[e.g.][]{Maoz&Badenes2010}, yet our models predict a small but nonzero CCSN contribution in this regime. Although CCSNe occurring after $400\,{\rm Myr}$ constitute only $\approx0.2\%$ of all CCSNe in this model, they appear across all of our model variations and are slightly more abundant at lower metallicities. Their presence, despite their rarity, indicates that very late time CCSN channels may exist and should not be dismissed when interpreting observed DTDs.

\begin{figure*}
    \centering
    \includegraphics[width=\textwidth]{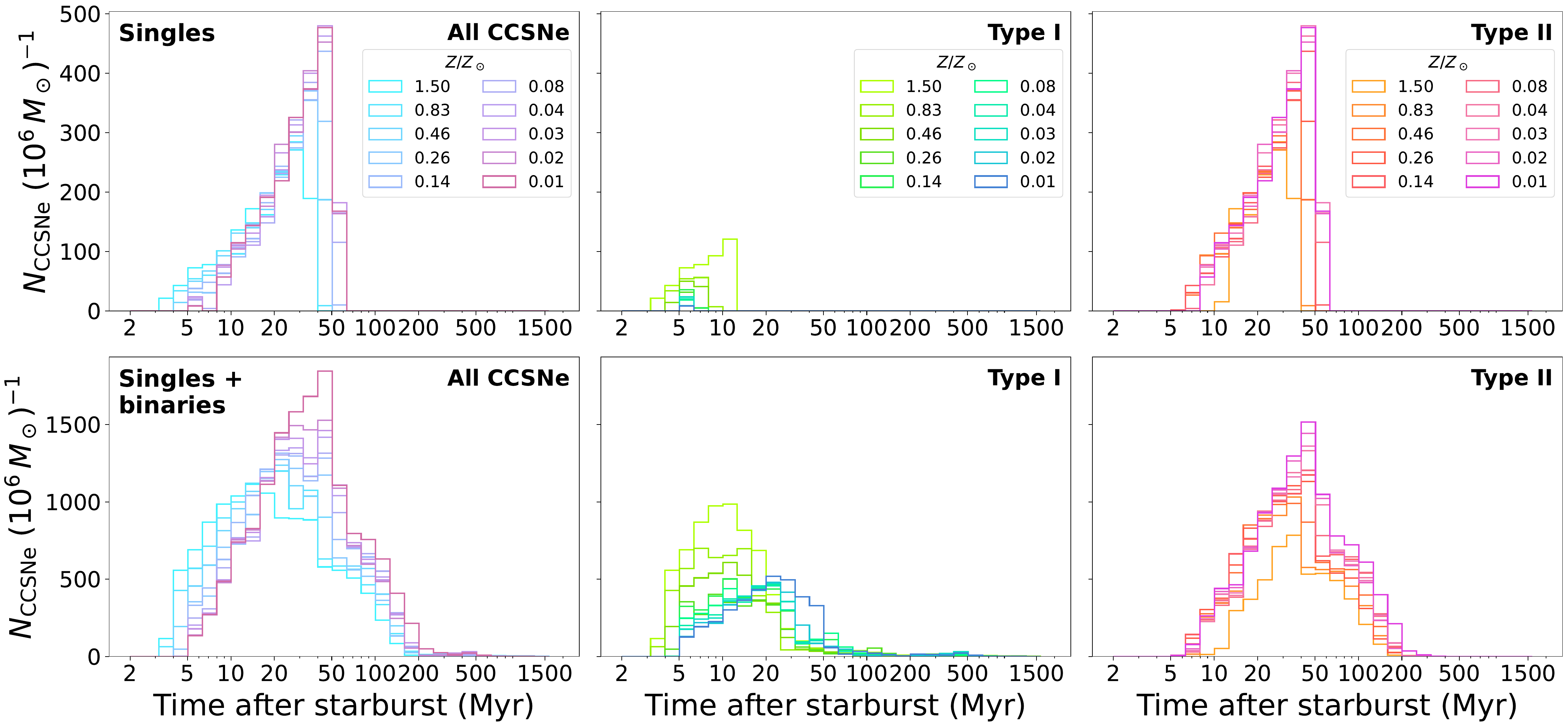}
    \caption{DTDs showing varied metallicity in single-only and binary-inclusive populations. The model shown follows our fiducial binary evolution assumptions ($\sigma=265\,{\rm km\,s}^{-1}$, $\alpha=1$) and spans ten metallicities evenly sampling our range. The three panels along the top show only single stars, while the lower panels include both single and binary systems. Columns correspond to all CCSNe (left), Type I only (middle), and Type II only (right). Only CCSNe forming remnants $M_{\rm rem}<15\,M_\odot$ are included.}
    \label{fig:DTDs_metallicity}
\end{figure*}

In Figure~\ref{fig:DTDs_metallicity} we present the DTDs of single stars, and the population including binaries across the range of metallicities we consider. We similarly present the entire population, and then divide the CCSNe into Type I, and then Type II.

Metallicity has two primary affects on the population of CCSNe. First, increasing metallicity increases the minimum mass for a single star to form a CCSN. At the lowest metallicity in our grid, the lowest mass single star which forms a CCSN in our model is $\approx6.8\,M_\odot$ and at our highest metallicity this rises to $\approx8.5M_\odot$. By increasing metallicity and thereby the minimum mass for a CCSN, we decrease the number of CCSNe in the population as a whole. We observe this trend in both our single and binary populations. As a result, lower metallicity delays and increases the peak of the DTD, since low mass stars are more abundant in the IMF. The second effect of metallicity is that it determines the minimum mass above which a single star strips its hydrogen envelope from winds. This translates to a changing ``transition time", before which all CCSNe from single stars are Type I, and after which all are Type II. As this time changes, the composition of the Type I and II populations change, but the overall DTD is unaffected. The impact is also present in the binary inclusive population, but interacting systems create a large overlap between Type I and II delay times.

\subsection{Binary Physics}

\begin{figure*}
    \centering
    \includegraphics[width=\textwidth]{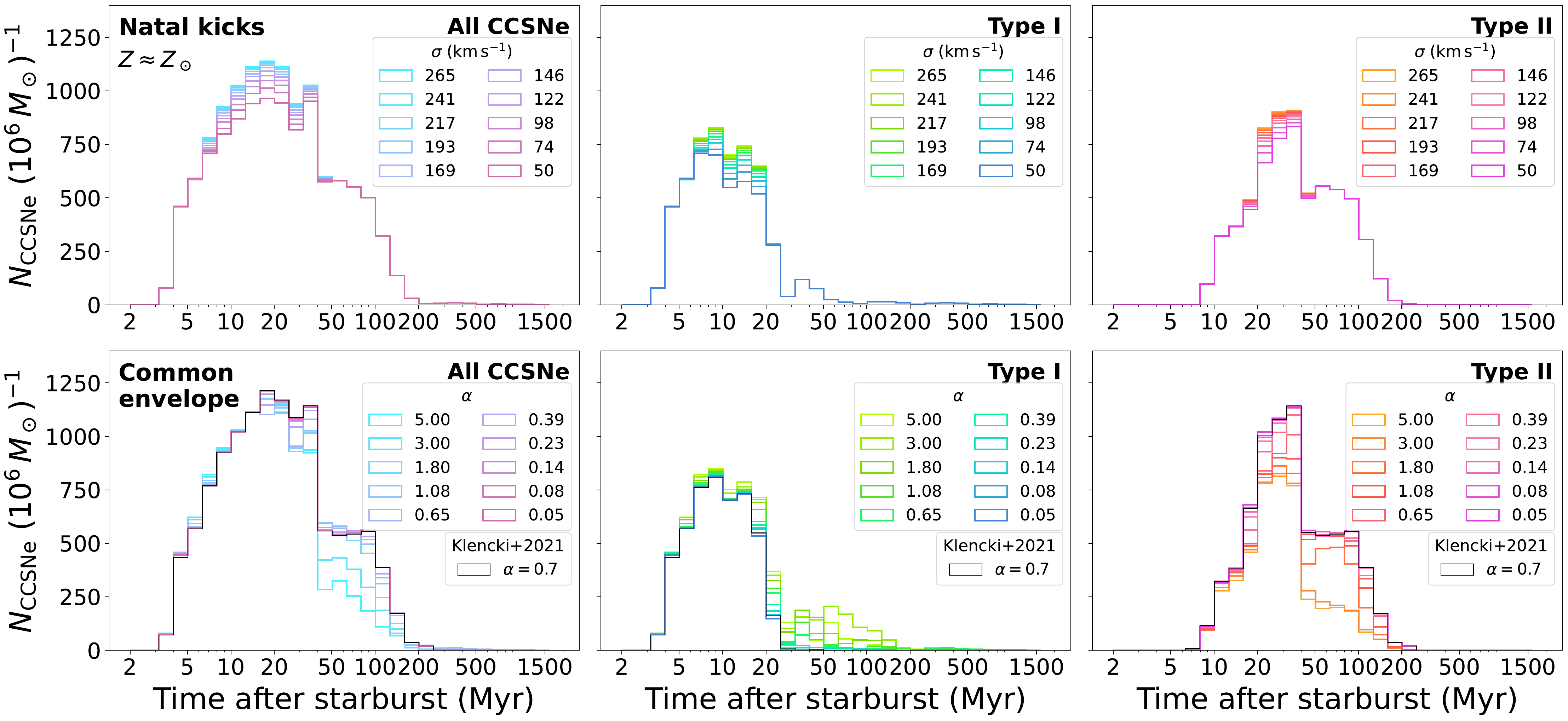}
    \caption{DTDs showing the effects of varying the natal kick dispersion and common-envelope efficiency. The models shown are computed at $Z\approx Z_\odot$. The top panels show models with different natal kick velocity dispersions ($\sigma$) across ten values spanning $50-265\,{\rm km\,s}^{-1}$. The lower panels show models with varying common-envelope efficiency ($\alpha$) across ten values spanning $0.05-5.0$. We also include our prescription based on the models of \citet{Klencki2021} with $\alpha=0.7$, shown as a black line in each lower panel. Columns correspond to all CCSNe (left), Type I only (middle), and Type II only (right). Only CCSNe forming remnants $M_{\rm rem}<15\,M_\odot$ are included.}
    \label{fig:DTDs_sigma_alpha}
\end{figure*}

The upper three panels of Figure~\ref{fig:DTDs_sigma_alpha} show the impact of varying the velocity dispersion for natal kicks ($\sigma$) on the total, Type I, and Type II population. Varying $\sigma$ only slightly changes the overall DTD shape. Larger natal kicks, which disrupt the binary, allow secondaries to produce CCSNe more easily. In the case where a binary remains bound after a natal kick, the resulting orbit is almost always eccentric. This increases the likelihood of mass transfer with the secondary as a donor. In the case where the secondary is massive enough to produce a CCSN (i.e. $\gtrsim8\,M_\odot$) and the primary is already a compact object, this mass transfer will lead to CEE and a subsequent merger which tidally disrupts the secondary.

When the natal kick disrupts the binary, the secondary undergoes single star evolution and will produce a CCSN if it is sufficiently massive. This will be Type I if the mass surpasses the ``transition mass" between Type I and Type II which was previously described for single stars, thus explaining why higher natal kicks lead to an increase in production of both Type I and Type II CCSNe.

The lower three panels of Figure~\ref{fig:DTDs_sigma_alpha} show populations with different $\alpha$, the CEE efficiency parameter, as well as the population with a custom prescription for CEE survival following the model in \citet{Klencki2021} with $\alpha=0.7$. Unlike our findings for modifying natal kick strengths, we find that varying our common envelope prescriptions strongly impacts the Type I and II CCSNe populations. Lowering $\alpha$ increases the frequency of stellar mergers from systems where neither star is sufficiently massive to explode independently, which greatly increases the proportion of late Type II CCSNe. Interestingly, we find that high $\alpha$ values generate a late time Type I CCSN population by enabling the survival of both phases of CEE in the white dwarf-helium star merger pathway described above. Thus, while raising $\alpha$ decreases the total number of late CCSNe and stellar mergers, this result is surprisingly negated by enabling these systems, which ultimately merge due to their short orbital periods.

In the $\lesssim15\,{\rm Myr}$ regime, we see a different impact of this effect. The systems with massive donors which would merge and produce a delayed CCSN with low $\alpha$ can successfully eject the shared envelope with $\alpha>1$. This leads to an early CCSN from the primary, thus increasing the production of CCSNe before $15\,\rm{Myr}.$ 

We find that the models which follow \citet{Klencki2021} tend to behave very similarly to our very low $\alpha$ models. A substantial portion of late CCSNe are produced, and these are overwhelmingly Type II. This model, which best represents the overall rate of Type I and Type II CCSNe across metallicities, does so by producing a substantial number of stellar mergers, and thereby suggesting that the most accurate DTDs to match CCSN type observations should have a substantial number of late CCSNe.

\subsection{CCSN Subtypes}
\label{DTD_subtypes}

Having discussed the delay times of Type I and II CCSNe broadly across models of binary evolution, we now turn to describing the DTDs of our specific subtypes within our preferred population following the model for CEE within \citet{Klencki2021}. We present the impact of our classifications on the resulting DTDs of subtypes in the appendix, in Figure \ref{fig:DTDs_classification}. However, we proceed here with our preferred classification scheme, restricting our analysis to solar metallicity, applying the \textit{``branch IIP first''} approach for our Type II CCSNe, and the \textit{``relative He''} scheme with $M_{\rm ej, He}/M_{\rm ej}=0.43$ for the Type I CCSNe. For each subtype, we present DTDs divided according to i) the ZAMS mass of the progenitor, ii) the mass at the time of core collapse, and iii) the mass of the resulting stellar remnant. The distributions of these quantities are also shown in the appendix, in Figure~\ref{fig:subtype_masses}.

\subsubsection{Type II DTDs}

\begin{figure*}
    \centering
    \includegraphics[width=\textwidth]{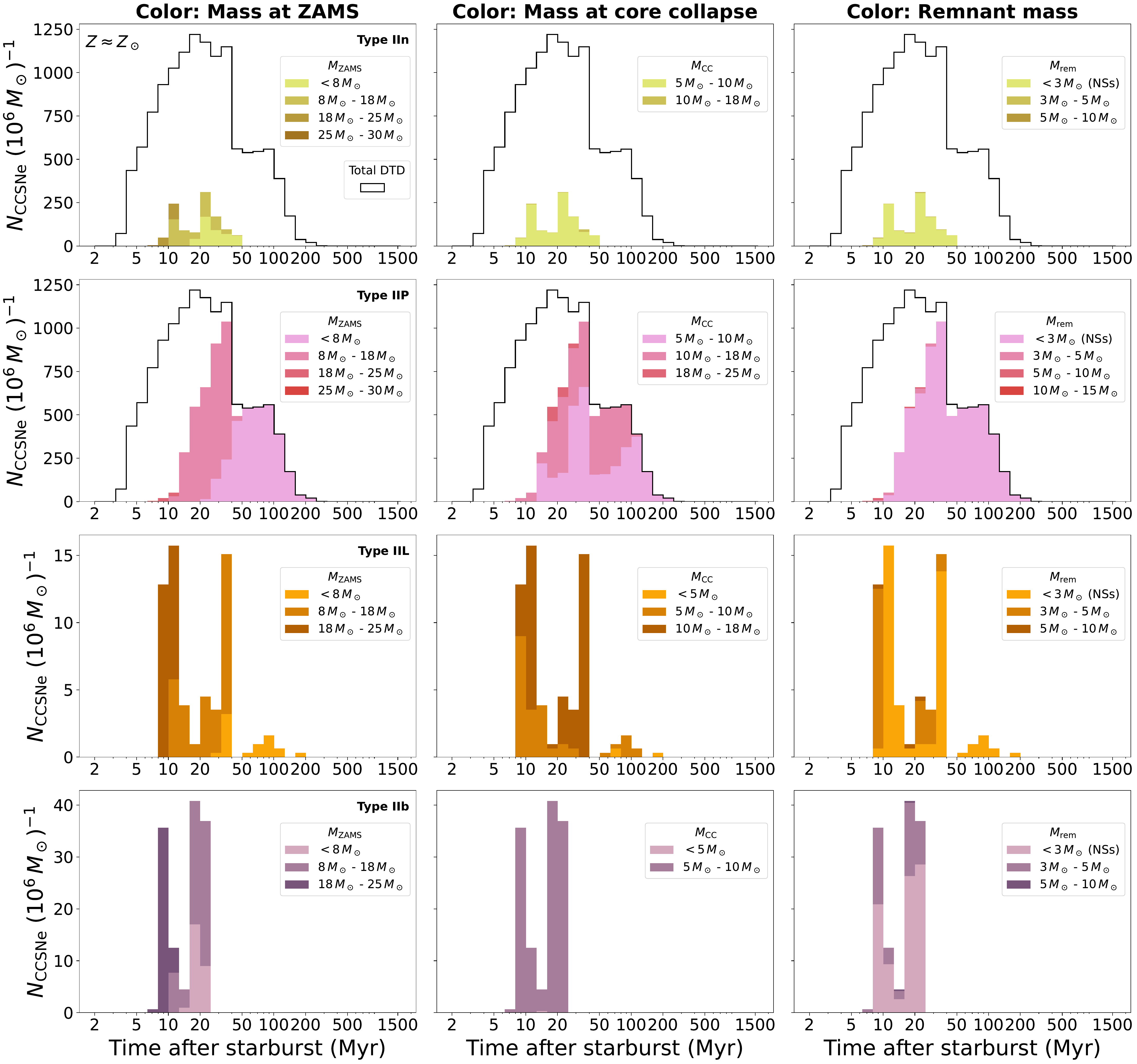}
    \caption{DTDs of Type II CCSN subtypes. The models shown use the CEE prescription based on \citet{Klencki2021} with $\alpha=0.7$ at $Z\approx Z_\odot$, and include only CCSNe forming remnants with $M_{\rm rem}<15\,M_\odot$. Rows correspond to subtypes IIn, IIP, IIL, and IIb (top to bottom). Columns show stacked histograms binned by initial mass at ZAMS (left), mass at core collapse (middle), and remnant mass (right). Stacked histograms separate contributions from the indicated mass ranges, and the total DTD is shown as a black outline where applicable. Total DTDs are not shown for Types IIL and IIb due to their small sample sizes, which would distort the scale.}
    \label{fig:DTDs_mass_II}
\end{figure*}

In Figure~\ref{fig:DTDs_mass_II} we present the subtype DTDs for our Type II CCSNe. We describe the important characteristics of each subtype below.

\smallskip \noindent \textbf{Type IIn:} We find that our IIn progenitors have a bimodal DTD, corresponding to a bimodal distribution of initial masses. The first peak around $11\,{\rm Myr}$ corresponds to more massive progenitors, typically $\sim16$-$20\,M_\odot$. The second peak around $23\,{\rm Myr}$ corresponds to progenitors with ZAMS masses $\sim5$-$9\,M_\odot$. We find that the final mass at core collapse regardless of initial mass, most typically is found to be nearly $\sim9$-$10\,M_\odot$, and most typically a NS around $\sim2.3M_\odot$ is formed.

\smallskip \noindent \textbf{Type IIP:} We find that the IIP population creates the vast majority of the CCSNe in our model after $\sim32\,{\rm Myr}$, and their DTD includes two distinct peaks. The first portion of their DTD up until $\sim38\,{\rm Myr}$ is dominated by progenitors with masses $>8\,M_\odot$, with the vast majority between $8$-$13\,M_\odot$, and a very small number extending up to $\sim24\,M_\odot$. The later portion of the DTD, which sustains a peak between $\sim50$-$100\,{\rm Myr}$ is made up entirely of progenitors not massive enough to form CCSNe as single stars (in our models $M_{\rm min,CCSN}\approx8.3\,M_\odot$ at $Z_\odot$). Most of these progenitors are $\sim3$-$6\,M_\odot$ at ZAMS. Of these late CCSNe, $\sim97\%$ originate from stellar merger progenitors, and the remaining $\sim3\%$ originate from low mass secondaries which accrete mass from a companion, causing them to be massive enough to explode. Despite the diversity in initial masses, the majority of the progenitors converge to $\sim7$-$11\,M_\odot$ at core collapse and produce NSs with masses typically less than $\sim1.7M_\odot$.

\smallskip \noindent \textbf{Type IIL:} The DTD of the Type IIL CCSNe falls into three primary regions. The later CCSNe with delay times $\gtrsim45\,{\rm Myr}$ all arise from stellar mergers, with progenitor stars with initial masses $<7\,M_\odot$. The majority of these mergers are from Hertzsprung Gap stars with oxygen-neon white dwarf companions. The CCSNe in the central region of the DTD, from $\sim20$-$45\,{\rm Myr}$ also originate from stellar merger progenitors. The merging stars are produced from a variety of evolutionary phases, but the most common are stars beginning core-helium burning and oxygen-neon white dwarfs. These progenitors have ZAMS masses from $\sim7$-$11\,{M_\odot}$. The early portion of the DTD, $\lesssim 20\,{\rm Myr}$, come from interacting binaries, but not stellar mergers. The progenitors have initial masses $\sim13$-$23\,M_\odot$, and nearly all undergo stable mass transfer with a companion, after which several solar masses of hydrogen still remain on the envelope. This mass transfer either begins at the onset of core helium burning, or on the early asymptotic giant branch. Most of these progenitors have masses of $\sim9$-$12\,M_\odot$ at core collapse, and nearly all form NSs ($M_{\rm rem}<3\,M_\odot$, but the rest form low mass BHs, all $\lesssim5.5\,M_\odot$.

\smallskip \noindent \textbf{Type IIb:} The first peak of the DTD around $\sim8$-$12\,{\rm Myr}$ originates from a combination of channels. Predominantly, these CCSNe form from non-interacting or single stars with initial masses $\sim20$-$24\,M_\odot$, which lose most of their envelope due to wind mass loss. The next most frequent channel is a failed CEE causing a merger between a donor crossing the Hertzsprung Gap or undergoing core helium burning and a MS companion. In this case, the donor has an initial mass $\sim16$-$24\,M_\odot$, and the accretor is much smaller, $\sim2$-$4\,M_\odot$, resulting in a merger product large enough to lose its envelope through increased winds. The final channel involved in this peak is includes systems with slightly smaller initial masses, but also extending to the same higher end, $\sim15$-$24\,M_\odot$, where a combination of winds and stable mass transfer removes most of the H envelope. 

The second peak around $\sim16$-$26\,{\rm Myr}$ originates predominately from systems with initial masses $\lesssim15\,M_\odot$ which lose the majority of their envelope from stable mass transfer. A few still are from MS-MS mergers with $M_1\approx M_2\approx10\,M_\odot$. Nearly all these progenitors are $\sim6.5$-$8\,M_\odot$ at core collapse, and produce NSs, with a few producing low-mass BHs, none more massive than $7\,M_\odot$. 

\subsubsection{Type I DTDs}

\begin{figure*}
    \centering
    \includegraphics[width=\textwidth]{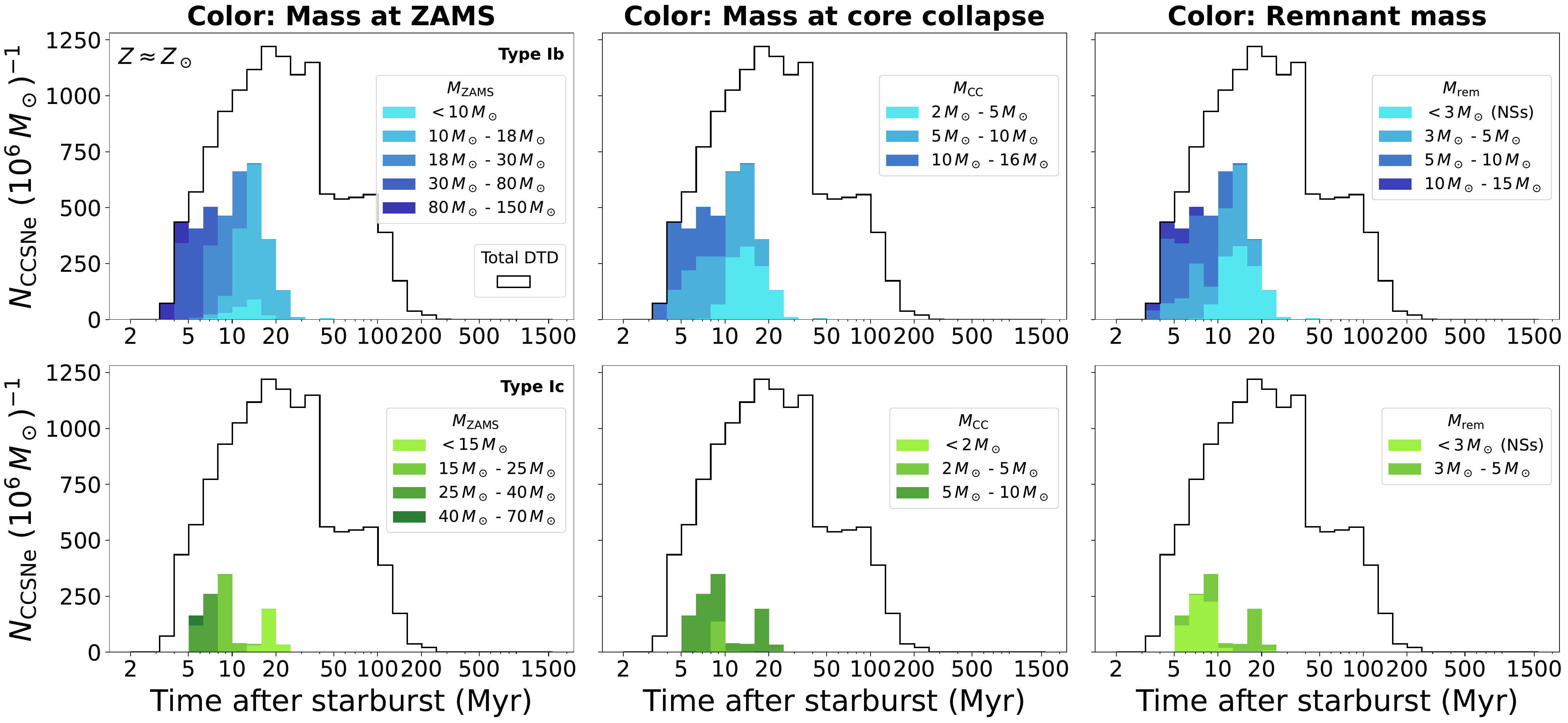}
    \caption{DTDs of Type I CCSN subtypes. The models shown use the CEE prescription based on \citet{Klencki2021} with $\alpha=0.7$ at $Z\approx Z_\odot$, and include only CCSNe forming remnants with $M_{\rm rem}<15\,M_\odot$. Rows correspond to subtypes Ib (top) and Ic (bottom). Columns show stacked histograms binned by initial mass at ZAMS (left), mass at core collapse (middle), and remnant mass (right). Stacked histograms separate contributions from the indicated mass ranges, and the total DTD is shown as a black outline.}
    \label{fig:DTDs_mass_I}
\end{figure*}

In Figure~\ref{fig:DTDs_mass_I} we show the DTDs of our Type Ib and Ic CCSNe. The progenitors of our Ib CCSNe originate from a much wider range of initial masses than our Ic population. The Ib initial masses are centered around a single wide peak from $\sim10$-$20\,M_\odot$, yet the upper tail extends to $\sim150\,M_\odot$. These high mass Ib progenitors make up the entirety of the DTD until $\sim4\,{\rm Myr}$, and Ib CCSNe dominate the DTD as the most abundant contribution for the first $\sim20\,{\rm Myr}$, peaking around $14\,{\rm Myr}$. At core collapse, the masses of these progenitors separate into two distinct non-overlapping populations, one from $\sim2$-$5\,M_\odot$, and the other from $\sim7.5$-$16\,M_\odot$. This lower mass group forms predominantly from progenitors with ZAMS masses $\sim10$-$20\,M_\odot$, nearly all of which lose their hydrogen envelope through stable mass transfer. The massive CCSNe progenitor population ($M>20\,M_{\odot}$), arises from a variety of progenitor channels. Roughly $20\%$ originate from non-interacting binaries or single stars, and have masses in the range $\sim21$-$105\,M_\odot$, though most commonly $\sim21$-$55\,M_\odot$. The remainder of the CCSNe originate from binaries which interact through stable or unstable mass transfer at some point before core collapse, yet they only donate some portion of their mass such that they still form massive helium stars. This low mass population at core collapse exclusively forms NSs, all $\lesssim1.8\,M_\odot$, whereas the higher mass population forms BHs, where the majority are $\lesssim8\,M_\odot$, but a small tail extends up to our limit of $15\,M_\odot$.

As noted in \citet{Eldridge2013}, our classification scheme leads to an overlapping continuum of Ib and Ic initial masses. Both the Ic DTD and the initial mass distribution are bimodal, with the two peaks straddling the lone peak of the Ib DTD and initial mass distribution. The first peak in the DTD lies near $\sim5$-$12\,{\rm Myr}$, and arises primarily from progenitors with masses $\sim16$-$30\,M_\odot$. These progenitors predominately lose their hydrogen envelope through stable mass transfer and lose some of their helium envelope through winds. The later peak in the DTD, from $\sim15$-$21\,{\rm Myr}$, corresponds to lower mass progenitors with masses between $\sim6$-$13\,M_\odot$, but a few extending up to $\sim17\,M_\odot$. These progenitors are predominately accretors that become sufficiently massive to remove most of their hydrogen envelope later through winds. These progenitors actually retain a few hundredths of a solar mass of their hydrogen envelope, so a lower threshold for hydrogen in Type IIb CCSNe might cause them to be identified as such. Despite this, they typically eject $\sim1.4$-$1.8\,M_\odot$ of helium, but also $\sim1.9$-$2.6\,M_\odot$ of carbon-oxygen material such that their ejecta composition still falls below the threshold. These two populations result in a bimodal mass distribution at core collapse, the first at $\sim4$-$5\,M_\odot$ and the second at $\sim7$-$9\,M_\odot$. The majority of these form NSs, but a few form low mass BHs with masses $\lesssim5\,M_\odot$.

\section{Conclusion} \label{sec:Conclusion}

We have performed a comprehensive population synthesis study of CCSNe using the rapid binary population synthesis code \cosmic. Across a broad range of metallicities, we explored the sensitivity of CCSN outcomes to variations in CEE, natal kicks, and remnant mass limited explodability. We developed physically motivated classification methods to separate CCSNe into Types I and II and their respective subtypes, and we presented the resulting subtype demographics and their DTDs. Our major results are as follows:

\smallskip \noindent \textbf{Binary Fraction:} Adopting the mass–dependent binary fraction of \citet{Offner23}, we find that about $95\%$ of CCSNe originate from binary progenitors, and that approximately $84\%$ of these systems interact before core collapse. This significantly enhances the predicted fraction of Type I CCSNe relative to single-star models, and highlights the need for accurate binary-interaction prescriptions to match observations.

\smallskip \noindent \textbf{Natal Kicks:} Increasing natal kick strength boosts the total number of CCSNe by enabling disrupted secondaries to evolve and explode without further binary interaction. This primarily increases the number of Type II CCSNe, although very massive accreting secondaries at high metallicity may instead evolve toward Type I CCSN outcomes. Lower kicks preferentially create eccentric post-SN binaries that undergo CEE and merge, suppressing second CCSN explosions. Natal kicks do not alter the \emph{timing} of secondary explosions, only whether they occur, modifying the amplitude but not the structure of the DTD. Natal kicks have a limited impact on the ratio of Type I/II CCSNe.

\smallskip \noindent \textbf{Common Envelope Evolution:} Varying the common envelope efficiency parameter $\alpha$ dramatically alters CCSN outcomes, but changing the model for mass transfer stability has only a modest impact. Reproducing the observed Type I/II CCSN ratios from supernova surveys \citep[e.g.,][]{KK12, LOSS} requires low $\alpha$, producing stellar mergers in nearly all CEE episodes. Alternatively, we can also match observations by implementing CEE prescriptions with envelope binding energy calculations based on detailed simulations following \citet{Klencki2021}. Higher $\alpha$ creates more hydrogen-rich merger products with delayed explosions, suggesting that DTDs of populations consistent with the observed Type I/II ratio should include a high fraction of late-time events. In addition, higher $\alpha$ more frequently yields hydrogen-poor merger products through channels that survive an initial CEE phase and merge later. Our models demonstrate that these pathways can produce very late Type I, hydrogen-poor merger product CCSNe, extending up to $\sim1550\,{\rm Myr}$, although such events are rare.

\smallskip \noindent \textbf{Type I vs. II Timing:} Type I CCSNe have systematically shorter delay times than Type II CCSNe across all metallicities because of their typically more massive progenitors evolving more quickly. Type I CCSNe yield both BH and NS remnants in comparable proportions, whereas Type II CCSNe predominantly form NSs, and do so almost exclusively at solar metallicity. Metallicity preserves the overall timing hierarchy between Types I and II, but shifts the characteristic times at which each type tends to occur. At high metallicity, stronger winds raise the minimum threshold mass required for a star to undergo core collapse, shifting the cumulative DTD to earlier times. High metallicities also lower the minimum mass above which a star strips its own hydrogen envelope through winds, effectively bringing the characteristic time for Type I CCSNe earlier. Finally, strong winds suppress iron–core formation in the most massive stars, allowing very massive progenitors to explode rather than collapse directly to black holes, which further increases early-time CCSNe. At low metallicity, weaker winds allow lower-mass stars to retain more of their envelopes and build larger cores, shifting the DTD toward later times and producing a pronounced late-time tail of Type II CCSNe. Metallicity therefore changes the location of the peak of the DTD substantially, from about $20\,{\rm Myr}$ at our highest metallicity to around $50\,{\rm Myr}$ at the lowest.

\smallskip \noindent \textbf{Remnant Mass Limit:} Enforcing a remnant mass threshold of $M_{\rm rem} \lesssim 3\,M_\odot$ for successful explosions is essential for reproducing the observed Type I/II CCSN ratio with our models. Without this restriction, high-metallicity populations substantially overproduce Type I CCSNe relative to observations. Although this assumption is not strictly physical, as many studies suggest that successful explosions should extend to higher remnant masses are not a monotonic function of progenitor mass \citep[e.g.][]{Sukhbold+2016:2016ApJ...821...38S, 2023ApJ...949...17B, Burrows+2025:2025ApJ...987..164B}, our results are tightly coupled to the adopted wind prescriptions, which strongly influence final core masses. Modulating the remnant mass limit monotonically does not significantly alter the subtype distribution among Type II CCSNe, but allowing higher remnant masses increases the predicted fraction of Type Ib CCSNe in our models.

\smallskip \noindent \textbf{CCSN Subtypes:}
Although we adopt physically motivated classification criteria, our models do not precisely reproduce the observed relative subtype proportions of Type II CCSNe, and most of the schemes we consider also fail to fully recover the observed Type I subtype ratios. Nevertheless, our simulations provide insight into the dominant formation channels of each subtype, and we ultimately identify a scheme that reproduces the observed Type I proportions.

\begin{itemize}

    \item \textit{IIn:} At low metallicity, pre-explosion mass loss in IIn progenitors arises from both wind mass loss and RLOF mass transfer, with mass transfer contributing $\approx 50\%$ of cases. This fraction declines to $\lesssim 3\%$ at solar metallicity, where winds dominate. We also find a substantial population of CCSNe with massive hydrogen envelopes that experience significant pre-core-collapse mass loss; whether these events are classified as IIn or IIP significantly affects subtype statistics.

    \item \textit{IIP:} These progenitors span the broadest range of initial masses and make up about $90\%$ of the Type II population in our models. More than half originate from stellar mergers, often through mass-ratio reversal: a secondary accretes the primary’s envelope, then later engulfs and merges with the primary during a CEE episode. These systems tend to explode at the latest times of any Type II subtype.

    \item \textit{IIL:} The fraction of IIL CCSNe declines steadily with metallicity. Their progenitors arise almost exclusively from binary interaction, with the dominant channels being mergers and stable mass transfer.

    \item \textit{IIb:} We significantly underproduce Type IIb CCSNe. This likely reflects limitations of COSMIC, which rarely leaves behind the thin, partially stripped hydrogen envelopes after RLOF necessary for IIb progenitors. Where they do appear, IIb progenitors tend to be among the most massive of the Type II population.

    \item \textit{Ib/c:} Type I CCSNe are highly sensitive to the classification scheme used. Using the strict helium ejecta mass threshold of $M_{\rm ej,He} = 0.14\,M_\odot$ we systematically underproduce Type Ic CCSNe. Classifying Ic progenitors as systems that donate mass while already helium stars increases the number of Ic events, but still falls short of observational constraints. This however highlights that up to $\sim 20\%$ of helium stars interact, suggesting that isolated helium-star evolution alone might be insufficient to explain the diversity of Ib/c progenitors.

    We achieve the best agreement with observations when separating Types Ib and Ic via the ratio of helium ejecta mass to total ejecta mass, finding that dividing on $M_{\rm ej,He}/M_{\rm ej} = 0.43$ reproduces the observed fraction of Ic CCSNe. This ratio remains robust even when adopting higher remnant mass limits, although increasing the remnant mass threshold does raise the proportion of Type Ib CCSNe.

    Across our models, Ib CCSNe typically arise from the most massive progenitors, yet Ib and Ic progenitors form an overlapping continuum of initial masses. At solar metallicity, Ic progenitors show a bimodal ZAMS-mass distribution, with the Ib distribution peaking between the two modes. Both Ib and Ic progenitors interact in $\gtrsim 95\%$ of cases, with hydrogen-envelope removal occurring predominantly through binary interaction rather than winds.

\end{itemize}

We have demonstrated that adopting a mass-dependent binary fraction strongly increases the predicted number of Type I CCSNe, that common envelope efficiency prescriptions dramatically alter CCSNe outcomes, that stellar mergers form an essential part of the CCSN population and must be incorporated to reproduce observations, and that mapping observational CCSNe subtypes to population synthesis outputs is complex but yields important insight into CCSNe formation channels. Several future directions are possible based on this work. In particular, application of hybrid population synthesis techniques like SEVN \citep{Iorio+2023:2023MNRAS.524..426I} or METISSE \citep{Agrawal+2020:2020MNRAS.497.4549A, Agrawal+2023:2023MNRAS.525..933A} will allow for more faithful comparison to predictions from population synthesis based on detailed binary evolution simulations like \citet{Souropanis2025} and \citet{Ercolino+2025:2025arXiv251004872E}. Hybrid population synthesis techniques, which provide access to detailed core structure from single star simulations, also enable more direct connections to detailed models for core collapse \citep{Patton+2020:2020MNRAS.499.2803P, Maltsev+2025:2025A&A...700A..20M}. Finally, as the CCSNe sample grows and calculations of volume- or magnitude-limited rates continue to be developed \citep[e.g.][]{Perley+2020:2020ApJ...904...35P}, direct comparisons between population synthesis calculations and measured rates may ultimately help to disentangle the rich landscape of stellar explosions.  

\begin{acknowledgments}
The authors thank Maria Drout, Ashley Villar, Anya Nugent, Adam Miller, Steve Schulze, Andrea Ercolino, and Dimitrios Souropanis for helpful conversations and editing suggestions. K.B. acknowledges partial support from the Falco-DeBenedetti Career Development Professorhip. M.M. acknowledges support from the Pennsylvania Space Grant Consortium and NSF Award \#2244348. A.O. acknowledges support from the McWilliams Postdoctoral Fellowship in the McWilliams Center for Cosmology and Astrophysics at Carnegie Mellon University.

This work made use of the following software packages: \texttt{astropy} \citep{astropy:2013, astropy:2018, astropy:2022}, \texttt{matplotlib} \citep{Hunter:2007}, \texttt{numpy} \citep{numpy}, \texttt{pandas} \citep{mckinney-proc-scipy-2010, pandas_10957263}, \texttt{python} \citep{python}, \texttt{scipy} \citep{2020SciPy-NMeth, scipy_12522488}, \texttt{COSMIC} \citep{Breivik2020, COSMIC_14828013}, \texttt{Cython} \citep{cython:2011}, \texttt{h5py} \citep{collette_python_hdf5_2014, h5py_4250762}, \texttt{schwimmbad} \citep{schwimmbad}, \texttt{seaborn} \citep{Waskom2021}, and \texttt{tqdm} \citep{tqdm_11107065}.

Software citation information aggregated using \texttt{\href{https://www.tomwagg.com/software-citation-station/}{The Software Citation Station}} \citep{software-citation-station-paper, software-citation-station-zenodo}.
\end{acknowledgments}

\clearpage
\appendix
\restartappendixnumbering
\invisiblesection{Appendix}

\begin{figure*}[h]
    \centering
    \includegraphics[width=\textwidth]{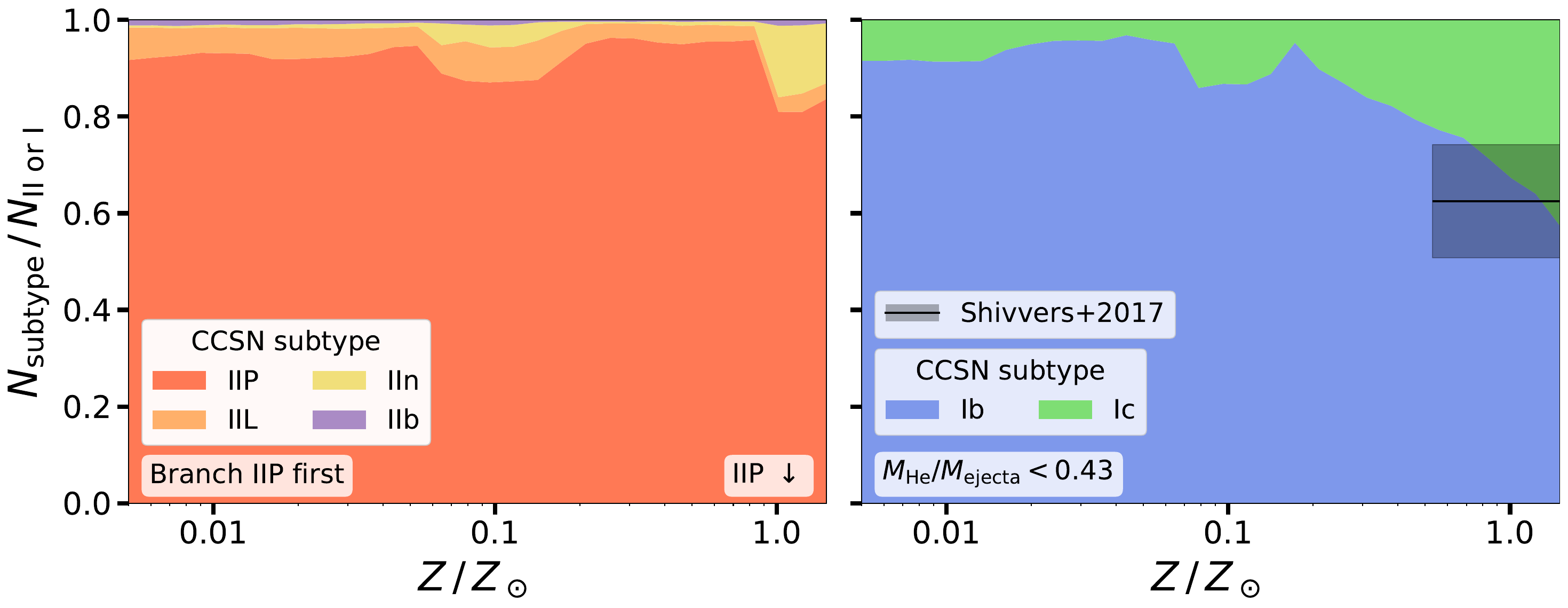}
    \caption{Stacked subtype fractions of Type II (left) and Type I (right) CCSNe across metallicity, with the alternative mass transfer stability prescription described in Section~\ref{Common Envelope Variations}. The model shown follows the CEE prescriptions of \citet{Klencki2021}, and only includes CCSNe forming small remnants ($M_{\rm rem}<3.0\,M_\odot$). The left panel applies the \textit{``branch IIP first''} classification (see Section~\ref{Classification method and variations II}) and the right panel applies the \textit{``relative helium''} scheme with $M_{\rm ej,He}/M_{\rm ej}=0.43$. Subtypes are colored as shown in the legend and stacked such that their sum is unity. The lower bound of the left panel is raised to aid in viewing since Type IIP CCSNe dominate the population.}
    \label{fig:subtypes_MT}
\end{figure*}

In this appendix, we present several supplementary figures that provide additional context and support for the results discussed in the main text. Figure~\ref{fig:subtypes_MT} illustrates the impact of adopting an alternative mass-transfer stability prescription (described in Section~\ref{Common Envelope Variations}). While this change does not significantly alter the overall Type I/II CCSN ratio or the shape of the DTDs, it does produce modest shifts in subtype demographics. Most notable is a slight increase in the fraction of Type IIL CCSNe around $Z/Z_\odot \approx 0.2$, which produces a smoother decline in the IIL rate with increasing metallicity. The other primary impact is a metallicity-wide increase in the proportion of Type Ib CCSNe. We still find that this mass transfer stability variation, when using the same $M_{\rm ej,He}/M_{\rm ej}=0.43$ threshold as in the main text, reproduces the Ib/Ic observations reported in \citet{Shivvers2017}.

\begin{figure*}
    \centering
    \includegraphics[width=\textwidth]{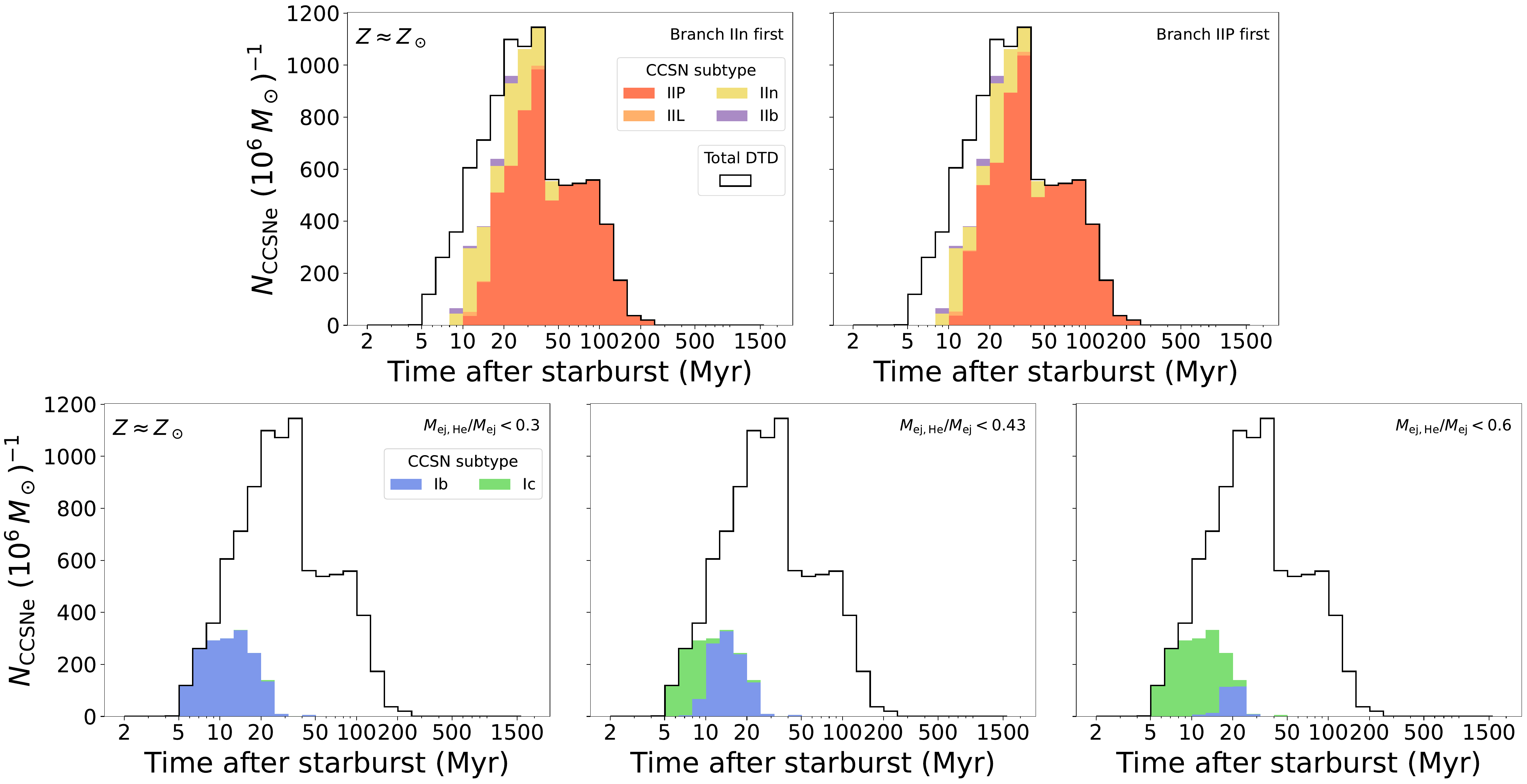}
    \caption{DTDs of CCSN subtypes with varied classification scheme criteria. The models shown use the CEE prescription based on \citet{Klencki2021} with $\alpha=0.7$ at $Z\approx Z_\odot$, and include only CCSNe forming remnants with $M_{\rm rem}<15\,M_\odot$. The top panels show all Type II subtypes stacked according to the classification scheme used: the \textit{``branch IIn first''} scheme (left) and the \textit{``branch IIP first''} scheme (right), see Section~\ref{Classification method and variations II} for the definition of these schemes. The lower panels show Type I CCSNe classified under the \textit{``relative helium''} scheme (see Section~\ref{Classification method and variations I}), with varying thresholds in the ratio of ejected helium to total ejecta mass, $M_{\rm ej,He}/M_{\rm ej}$ = 0.3 (left), 0.43 (middle), and 0.6 (right). Stacked histograms separate subtype contributions, and the total DTD is shown as a black outline in each panel.}
    \label{fig:DTDs_classification}
\end{figure*}

Figure~\ref{fig:DTDs_classification} shows how variations in our classification schemes influence the DTDs of both Type II and Type I CCSNe. The differences between the \textit{``branch IIn first''} and \textit{``branch IIP first''} schemes are minimal at solar metallicity, though the latter reclassifies some late-time IIn events as IIPs. Varying the ratio $M_{\rm ej,He}/M_{\rm ej}$ used to distinguish Ib and Ic CCSNe blends the Type I DTDs to differing degrees, increasing this ratio pushes Ic CCSNe toward later delay times.

\begin{figure*}[t]
    \centering
    \includegraphics[width=\textwidth]{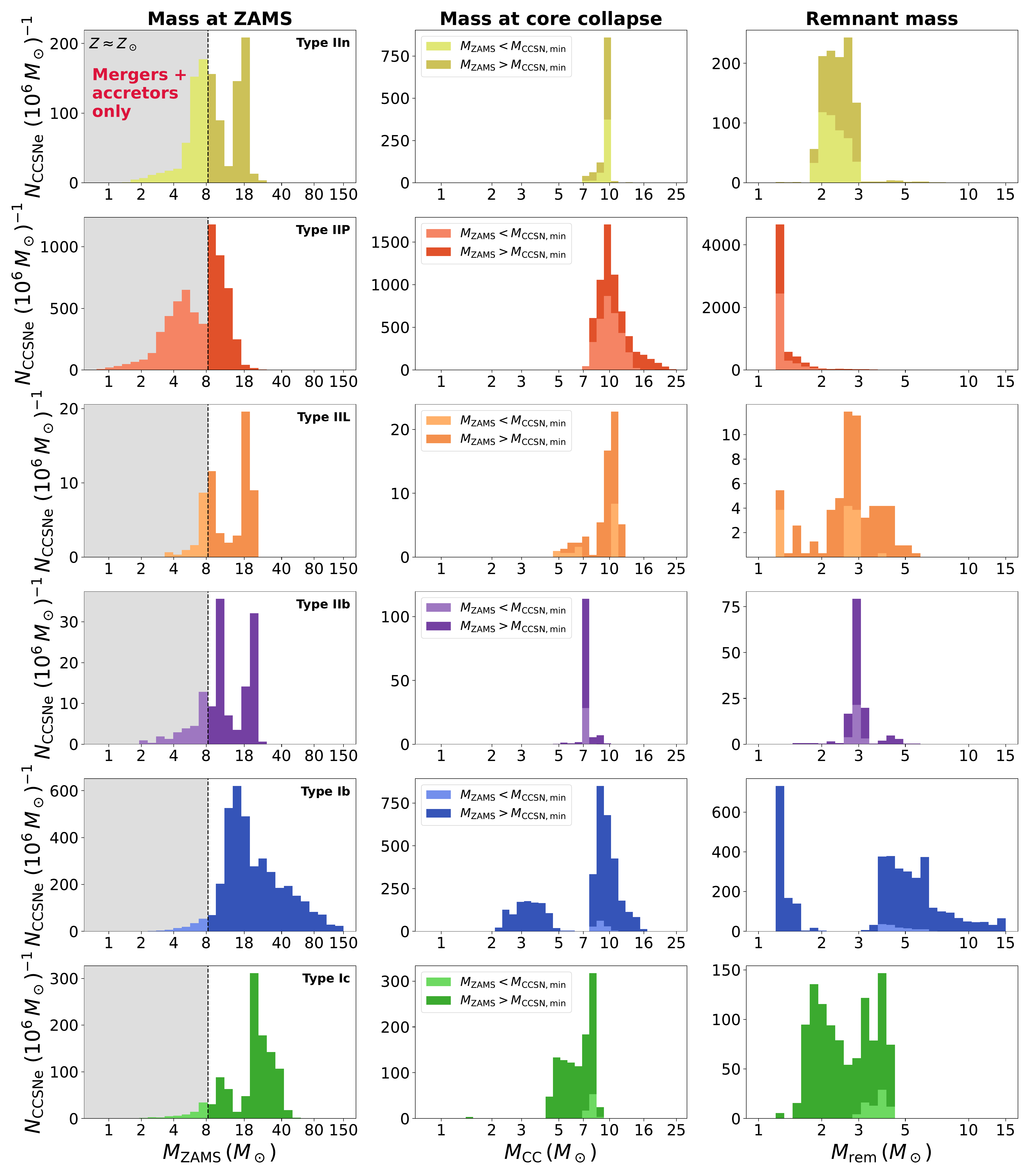}
    \caption{Distributions of ZAMS, core collapse, and remnant masses for all CCSN subtypes. The models shown use the CEE prescription based on \citet{Klencki2021} with $\alpha=0.7$ at $Z\approx Z_\odot$, and include only CCSNe forming remnants with $M_{\rm rem}<15\,M_\odot$. Columns show the distributions of initial mass at ZAMS (left), mass at core collapse (middle), and remnant mass (right). The y-axis in all panels shows counts normalized by the mass of the initial sample.}
    \label{fig:subtype_masses}
\end{figure*}

Finally, Figure~\ref{fig:subtype_masses} summarizes the distributions of ZAMS, core collapse, and remnant masses for all CCSN subtypes. These distributions provide useful context for interpreting the stacked DTDs shown in Figures~\ref{fig:DTDs_mass_II} and~\ref{fig:DTDs_mass_I}, clarifying the mass ranges of different progenitor channels. This figure is used for analysis in Sec~\ref{DTD_subtypes}.

\clearpage

\bibliography{sample701}{}
\bibliographystyle{aasjournalv7}

\end{document}